\documentclass{elsart}
\usepackage{graphicx} 
\journal{Parallel and Distributed Computing}

\begin{document}                  
\newcommand{\mytheoremcounter}{section}
\newcommand{\cheapHack}{ }
\newcommand{\BAPrn}{$\mathrm{BAP}_r$}
\newcommand{\BAPr}{$\mathrm{BAP}_r$\ }
\newcommand{\BAPsn}{$\mathrm{BAP}_s$}
\newcommand{\BAPs}{$\mathrm{BAP}_s$\ }
\newcommand{\BAPsrn}{$\mathrm{BAP}_{sr}$}
\newcommand{\BAPsr}{$\mathrm{BAP}_{sr}$\ }
\newcommand{\BAPucn}{$\mathrm{BAP}_{sr}$}
\newcommand{\BAPuc}{$\mathrm{BAP}_{sr}$\ }
\newcommand{\BSPrn}{$\mathrm{BSP}_r$}
\newcommand{\BSPr}{$\mathrm{BSP}_r$\ }
\newcommand{\BSPsn}{$\mathrm{BSP}_s$}
\newcommand{\BSPs}{$\mathrm{BSP}_s$\ }
\newcommand{\BSPsrn}{$\mathrm{BSP}_{sr}$}
\newcommand{\BSPsr}{$\mathrm{BSP}_{sr}$\ }
\newcommand{\BSPucn}{$\mathrm{BSP}_{sr}$}
\newcommand{\BSPuc}{$\mathrm{BSP}_{sr}$\ }
\newcommand{\NBAPrn}{$\mathrm{NBAP}_r$}
\newcommand{\NBAPr}{$\mathrm{NBAP}_r$\ }
\newcommand{\NBAPsn}{$\mathrm{NBAP}_s$}
\newcommand{\NBAPs}{$\mathrm{NBAP}_s$\ }
\newcommand{\NBAPsrn}{$\mathrm{NBAP}_{sr}$}
\newcommand{\NBAPsr}{$\mathrm{NBAP}_{sr}$\ }
\newcommand{\NBAPucn}{$\mathrm{NBAP}_{sr}$}
\newcommand{\NBAPuc}{$\mathrm{NBAP}_{sr}$\ }
\newcommand{\fixme}[1]{$\spadesuit$\marginpar{\tiny$\spadesuit$#1}}
\newcommand{\Xhalf}{X_\frac{1}{2}}
\newcommand{\Shalf}{\cS_{\frac{n}{2},n}}
\newcommand{\chih}{\chi_n}
\newcommand{\notchih}{\overline{\chi_n}}
\newcommand{\upsh}{\upsilon_n}
\newcommand{\notupsh}{\overline{\upsilon_n}}
\newcommand{\singlespacing}{\baselineskip 1em}
\newcommand{\onehalfspacing}{\baselineskip 1.25em}
\newcommand{\doublespacing}{\baselineskip 1.75em}
\newcommand{\truedoublespacing}{\baselineskip 2em}
\newcommand{\normalspacing}{\singlespacing}
\newcommand{\Paccept}[1]{{\Pr[#1 = \mathrm{accept}]}}
\newcommand{\Maj}{\mathit{Maj}}
\newcommand{\Nand}{\uparrow}
\newcommand{\Nor}{\downarrow}
\newcommand{\myem}[1]{{\bf #1}}
\newcommand{\reccr}[1]{\overline{\nu}(#1)}
\newcommand{\regcr}[1]{\nu(#1)}
\newcommand{\LOGDCFLclass}{\mathbf{LOGDCFL}}
\newcommand{\NLclass}{\mathbf{NL}}
\newcommand{\ACclass}{\mathbf{AC}}
\newcommand{\NCclass}{\mathbf{NC}}
\newcommand{\SCclass}{\mathbf{SC}}
\newcommand{\RCclass}{\mathbf{RC}}
\newcommand{\coNLclass}{\mathbf{co\!-\!NL}}
\newcommand{\Lclass}{\mathbf{L}}
\newcommand{\Lpoly}{\mathbf{L}_{/\mathrm{poly}}}
\newcommand{\Pclass}{\mathbf{P}}
\newcommand{\BPclass}{\mathbf{P}_{BP}}
\newcommand{\BPwidth}[1]{\mathbf{P}_{BP}^{#1}}
\newcommand{\NPclass}{\mathbf{NP}}
\newcommand{\coNPclass}{\mathbf{coNP}}
\newcommand{\PPclass}{\mathbf{PP}}
\newcommand{\BPPclass}{\mathbf{BPP}}
\newcommand{\ZPPclass}{\mathbf{ZPP}}
\newcommand{\RPclass}{\mathbf{RP}}
\newcommand{\coRPclass}{\mathbf{co\!-\!RP}}
\newcommand{\SigmaP}[1]{\mathbf{\Sigma_{#1}P}}
\newcommand{\PHclass}{\mathbf{PH}}
\newcommand{\PSPACE}{\mathbf{PSPACE}}
\newcommand{\RSPACE}{\mathbf{RSPACE}}
\newcommand{\DSPACE}{\mathbf{DSPACE}}
\newcommand{\imin}{{\mathit{min}}}
\newcommand{\imax}{{\mathit{max}}}
\newcommand{\gbf}{{\mathbf{g}}}
\newcommand{\zbf}{{\mathbf{z}}}
\newcommand{\wbf}{{\mathbf{w}}}
\newcommand{\vbf}{{\mathbf{v}}}
\newcommand{\xbf}{{\mathbf{v}}}
\newcommand{\onebf}{{\mathbf{1}}}
\newcommand{\zerobf}{{\mathbf{0}}}
\newcommand{\uvec}{\vec{u}}
\newcommand{\vvec}{\vec{v}}
\newcommand{\xvec}{\vec{x}}
\newcommand{\Deltap}{\Delta^\prime}
\newcommand{\Sigmap}{{\Sigma^\prime}}
\newcommand{\alphap}{\alpha^\prime}
\newcommand{\betap}{\beta^\prime}
\newcommand{\betapp}{\beta^{\prime\prime}}
\newcommand{\gammap}{\gamma^\prime}
\newcommand{\mup}{\mu^\prime}
\newcommand{\mupp}{\mu^{\prime\prime}}
\newcommand{\nubar}{{\overline{\nu}}}
\newcommand{\nubars}{\nubar^*}
\newcommand{\nus}{\nu^*}
\newcommand{\betas}{\beta^*}
\newcommand{\deltas}{\delta^*}
\newcommand{\deltap}{\delta^\prime}
\newcommand{\deltapp}{\delta^{\prime\prime}}
\newcommand{\lambdabar}{\overline{\lambda}}
\newcommand{\lambdap}{\lambda^\prime}
\newcommand{\lambdapp}{\lambda^{\prime\prime}}
\newcommand{\pip}{\pi^\prime}
\newcommand{\pipp}{{\pi^{\prime\prime}}}
\newcommand{\Psip}{\Psi^\prime}
\newcommand{\psip}{\psi^\prime}
\newcommand{\phip}{\phi^\prime}
\newcommand{\sigmap}{\sigma^\prime}
\newcommand{\etap}{\eta^\prime}
\newcommand{\etapp}{{\eta^{\prime\prime}}}
\newcommand{\epsilonp}{{\epsilon^\prime}}
\newcommand{\epsilonpp}{\epsilon^{\prime\prime}}
\newcommand{\epsilonppp}{\epsilon^{\prime\prime\prime}}
\newcommand{\ap}{a^\prime}
\newcommand{\bp}{b^\prime}
\newcommand{\cp}{c^\prime}
\newcommand{\up}{u^\prime}
\newcommand{\fp}{f^\prime}
\newcommand{\ep}{e^\prime}
\newcommand{\gp}{g^\prime}
\newcommand{\ip}{i^\prime}
\newcommand{\jp}{j^\prime}
\newcommand{\mpr}{m^\prime}
\newcommand{\np}{n^\prime}
\newcommand{\op}{o^\prime}
\newcommand{\qp}{q^\prime}
\newcommand{\rp}{r^\prime}
\newcommand{\spr}{s^\prime}
\newcommand{\tp}{t^\prime}
\newcommand{\gpp}{g^{\prime\prime}}
\newcommand{\ipp}{i^{\prime\prime}}
\newcommand{\jpp}{j^{\prime\prime}}
\newcommand{\epp}{e^{\prime\prime}}
\newcommand{\qpp}{q^{\prime\prime}}
\newcommand{\rpp}{r^{\prime\prime}}
\newcommand{\vp}{v^\prime}
\newcommand{\ypp}{y^{\prime\prime}}
\newcommand{\xpp}{x^{\prime\prime}}
\newcommand{\zp}{z^\prime}
\newcommand{\hp}{{h^\prime}}
\newcommand{\lp}{{l^\prime}}
\newcommand{\zpp}{{z^{\prime\prime}}}
\newcommand{\kp}{{k^\prime}}
\newcommand{\Dp}{{D^\prime}}
\newcommand{\Pp}{P^\prime}
\newcommand{\Ppp}{P^{\prime\prime}}
\newcommand{\Qp}{Q^\prime}
\newcommand{\Qpp}{Q^{\prime\prime}}
\newcommand{\Spr}{S^\prime}
\newcommand{\Tp}{T^\prime}
\newcommand{\Ep}{E^\prime}
\newcommand{\Dbar}{\overline{D}}
\newcommand{\Lp}{L^\prime}
\newcommand{\Lpp}{L^{\prime\prime}}
\newcommand{\Lbar}{\overline{L}}
\newcommand{\Lhat}{\widehat{L}}
\newcommand{\Ltilde}{\tilde{L}}
\newcommand{\Lcap}{{L^\cap}}
\newcommand{\Lcup}{{L^\cup}}
\newcommand{\Lpbar}{\overline{\Lp}}
\newcommand{\Mp}{M^\prime}
\newcommand{\Mpp}{M^{\prime\prime}}
\newcommand{\Mbar}{\overline{M}}
\newcommand{\Np}{N^\prime}
\newcommand{\Npp}{N^{\prime\prime}}
\newcommand{\Rp}{R^\prime}
\newcommand{\xbar}{\bar{x}}
\newcommand{\xp}{x^\prime}
\newcommand{\yp}{y^\prime}
\newcommand{\Uhat}{\widehat{U}}
\newcommand{\Up}{U^\prime}
\newcommand{\Upp}{U^{\prime\prime}}
\newcommand{\Vp}{V^\prime}
\newcommand{\Vhat}{\widehat{V}}
\newcommand{\Vbar}{\overline{V}}
\newcommand{\Ap}{{A^\prime}}
\newcommand{\App}{{A^{\prime\prime}}}
\newcommand{\Cp}{C^\prime}
\newcommand{\Fp}{F^\prime}
\newcommand{\Gp}{G^\prime}
\newcommand{\Gtilde}{\tilde{G}}
\newcommand{\Fpp}{F^{\prime\prime}}
\newcommand{\Zf}{{\mathbb{Z}}}
\newcommand{\Qf}{{\mathbb{Q}}}
\newcommand{\Rf}{{\mathbb{R}}}
\newcommand{\Cf}{{\mathbb{C}}}
\providecommand{\mathbb}[1]{\Bbb{#1}}
\newcommand{\qacc}{q^{acc}}
\newcommand{\qrej}{q^{rej}}
\newcommand{\qnon}{q^{non}}
\newcommand{\Qadd}{Q_{add}}
\newcommand{\Qacc}{Q_{acc}}
\newcommand{\Qrej}{Q_{rej}}
\newcommand{\Qnon}{Q_{non}}
\newcommand{\Qhalt}{Q_{halt}}
\newcommand{\Qjunk}{Q_{junk}}
\newcommand{\Qaccp}{{Q_{acc}^\prime}}
\newcommand{\Qrejp}{{Q_{rej}^\prime}}
\newcommand{\Qnonp}{{Q_{non}^\prime}}
\newcommand{\Qhaltp}{{Q_{halt}^\prime}}
\newcommand{\Qjunkp}{{Q_{junk}^\prime}}
\newcommand{\Qaccpp}{{Q_{acc}^{\prime\prime}}}
\newcommand{\Qrejpp}{{Q_{rej}^{\prime\prime}}}
\newcommand{\Qnonpp}{{Q_{non}^{\prime\prime}}}
\newcommand{\Qjunkpp}{{Q_{junk}^{\prime\prime}}}
\newcommand{\Cacc}{C_{acc}}
\newcommand{\Crej}{C_{rej}}
\newcommand{\Cnon}{C_{non}}
\newcommand{\Eacc}{E_{acc}}
\newcommand{\Erej}{E_{rej}}
\newcommand{\Enon}{E_{non}}
\def\cent{{\hbox{\rm\rlap/c}}}
\newcommand{\centp}{{\cent}^\prime}
\newcommand{\Bra}[1]{{\langle{#1}|}}
\newcommand{\Ket}[1]{{|{#1}\rangle}}
\newcommand{\BraKet}[2]{{\langle{#1}|{#2}\rangle}}
\newcommand{\iprod}[2]{{\langle{#1},{#2}\rangle}}
\newtheorem{theorem}{{\bf Theorem}}[\mytheoremcounter]
\newtheorem{lemma}[theorem]{{\bf Lemma}}
\newtheorem{aside}[theorem]{{\bf Aside}}
\newtheorem{example}[theorem]{{\bf Example}}
\newtheorem{question}[theorem]{{\bf Question}}
\newtheorem{answer}[theorem]{{\bf Answer}}
\newtheorem{conjecture}[theorem]{{\bf Conjecture}}
\newtheorem{proposition}[theorem]{{\bf Proposition}}
\newtheorem{property}[theorem]{{\bf Property}}
\newtheorem{corollary}[theorem]{{\bf Corollary}}
\newtheorem{observation}[theorem]{{\bf Observation}}
\newtheorem{definition}[theorem]{{\bf Definition}}
\newtheorem{remark}[theorem]{{\bf Remark}}
\newtheorem{thoughts}[theorem]{{\bf Thoughts}}
\newenvironment{proof}{ \begin{trivlist} 
                        \item \vspace{-\topsep} \noindent{\bf Proof:}\ }
                      {\rule{5pt}{5pt}\end{trivlist}}
\newcommand{\Case}[2]{\noindent{\bf Case #1:}#2}
\newcommand{\Subcase}[2]{\noindent{\bf Subcase #1:}#2}
\newcommand{\Half}{\frac{1}{2}}
\newcommand{\RtHalf}{\frac{1}{\sqrt{2}}}
\newcommand{\cA}{{\mathcal{A}}}
\newcommand{\cC}{{\mathcal{C}}}
\newcommand{\cE}{{\mathcal{E}}}
\newcommand{\cF}{{\mathcal{F}}}
\newcommand{\cH}{{\mathcal{H}}}
\newcommand{\cI}{{\mathcal{I}}}
\newcommand{\cK}{{\mathcal{K}}}
\newcommand{\cL}{{\mathcal{L}}}
\newcommand{\cM}{{\mathcal{M}}}
\newcommand{\cO}{{\mathcal{O}}}
\newcommand{\cP}{{\mathcal{P}}}
\newcommand{\cR}{{\mathcal{R}}}
\newcommand{\cS}{{\mathcal{S}}}
\newcommand{\cU}{{\mathcal{U}}}
\newcommand{\Span}{{\mathit{Span}}}
\newcommand{\Ch}[2]{{#1 \choose #2}}
\newcommand{\Ul}[1]{{\underline{#1}}}
\newcommand{\Floor}[1]{{\lfloor #1 \rfloor}}
\newcommand{\ignore}[1]{}
\newcommand{\noignore}[1]{#1}

\newcommand{\RMO}{\mathbf{RMO}}
\newcommand{\UMO}{\mathbf{UMO}}
\newcommand{\RMOe}{\mathbf{RMO}_\epsilon}
\newcommand{\RMM}{\mathbf{RMM}}
\newcommand{\UMM}{\mathbf{UMM}}
\newcommand{\RMMe}{\mathbf{RMM}_\epsilon}

\newcommand{\MOQFA}{\mathbf{MOQFA}}
\newcommand{\MOQFAe}{\mathbf{MOQFA}_\epsilon}
\newcommand{\MMQFA}{\mathbf{MMQFA}}
\newcommand{\MMQFAe}{\mathbf{MMQFA}_\epsilon}
\newcommand{\GQFA}{\mathbf{GQFA}}
\newcommand{\GQFAe}{\mathbf{GQFA}_\epsilon}

\newcommand{\REG}{\mathbf{REG}}
\newcommand{\PFA}{\mathbf{PFA}}
\newcommand{\PFAe}{\mathbf{PFA}_\epsilon}
\newcommand{\GFA}{\mathbf{GFA}}

% Code environment
\newcommand{\Foreach}[2]{\\{\bf\tt{for\ each}} $#1$ {\bf\tt{do}}\+ #2
\- \\ {\bf\tt{rof}}}
\newcommand{\Forloop}[2]{\\{\bf\tt{for}} $#1$ {\bf\tt{do}}\+ #2
\- \\ {\bf\tt{rof}}}
\newcommand{\Ifthen}[2]{\\{\bf\tt{if}} $#1$ {\bf\tt{then}}\+ #2
\- \\ {\bf\tt{fi}}}
\newcommand{\Ifelse}[3]{\\{\bf\tt{if}} $#1$ {\bf\tt{then}}\+ #2
\- \\ {\bf\tt{else}}\+ #3 \- \\ {\bf\tt{fi}}}
\newcommand{\Stmt}[1]{\\$#1$;}
\newcommand{\StartStmt}[1]{\+\kill$#1$;}
\newenvironment{pseudocode}{\begin{tabbing} 
\ \ \ \ \=\ \ \ \ \=\ \ \ \ \=\ \ \ \ \=\ \ \ \ \=\ \ \ \ \=\ \ \ \ \=\
\ \ \ \= } {\end{tabbing}}

% Moving proofs around
% #1 = proof \name   #2 = appendix ref    #3 = proof
\providecommand{\SaveProof}[3]{#3}
% #1 = proof \name   #2 = appendix ref    #3 = proof    #4 = sketch
\providecommand{\SketchProof}[4]{#3}
% #1 = proof \name   #2 = appendix label  #3 = title
\providecommand{\AppendixProof}[3]{}

\newcommand{\MoveProofsToAppendix}{\include{movemacs}}

% Short equation separator
\newcommand{\ShortSep}{\\ & &}
\newcommand{\LongSep}{}
\providecommand{\DefSep}{\LongSep}
\newcommand{\UseShortSep}{\renewcommand{\DefSep}{\ShortSep}}

% select abstract mechanism
\newcommand{\UseAbstract}[2]{#1}
\newcommand{\UseOtherAbstract}{\include{absselect}}

% EVIL stuff to make it fit for 10 page limit (two lines per page extra)
\newcommand{\StretchPage}{ \addtolength{\textheight}{0.05\textheight}
                           \addtolength{\topmargin}{-0.03\textheight}
                         }

% figure macro

\newcommand{\DoFigure}[4]{
                          \begin{figure}[ht]
\setlength{\topsep}{-10pt}
\setlength{\parsep}{0pt}
\setlength{\partopsep}{0pt}
\setlength{\parskip}{0pt}
                            \begin{center}
                              \ \includegraphics[scale=#2]{#1}\ 
                            \end{center}
                            \caption{#3\label{#4}}
                          \end{figure}
                         }

\newcommand{\DoBiFigure}[5]{
                          \begin{figure}[ht]
\setlength{\topsep}{-10pt}
\setlength{\parsep}{0pt}
\setlength{\partopsep}{0pt}
\setlength{\parskip}{0pt}
                            \begin{center}
                              \mbox{\ \includegraphics[scale=#3]{#1}\ 
                                    \hspace{1.0in}
                                    \ \includegraphics[scale=#3]{#2}\ }
                            \end{center}
                            \caption{#4\label{#5}}
                          \end{figure}
                         }

\newcommand{\DoDiFigure}[8]{ 
                          \begin{figure}[ht]
\setlength{\topsep}{-10pt}
\setlength{\parsep}{0pt}
\setlength{\partopsep}{0pt}
\setlength{\parskip}{0pt}
                            \begin{center}
                              \begin{minipage}[b]{0.35\linewidth}
                                \begin{center}
                                  \includegraphics[scale=#2]{#1}
                                \end{center}
                                \caption{#3\label{#4}}
                              \end{minipage}
                              \hspace{1.0in}
                              \begin{minipage}[b]{0.35\linewidth}
                                \begin{center}
                                  \includegraphics[scale=#6]{#5}
                                \end{center}
                                \caption{#7\label{#8}}
                              \end{minipage}
                            \end{center}
                          \end{figure}
                         }

\newcommand{\DoTable}[3]{
                          \begin{table}[ht]
                            \begin{center}
                              #1
                            \end{center}
                            \caption{#2\label{#3}}
                          \end{table}
                         }

\newcommand{\captionfontstyle}{\bf\footnotesize}

\begin{frontmatter}
\title{On the Complexity of Buffer Allocation in Message Passing Systems}

\author{Alex Brodsky},
\ead{abrodsky@cs.ubc.ca}
\author{Jan B\ae{}kgaard Pedersen},
\ead{matt@cs.ubc.ca}
\author{Alan Wagner}
\ead{wagner@cs.ubc.ca}
\address{Department of Computer Science,\\ 
University of British Columbia,\\ 
201-2366 Main Mall,\\ 
Vancouver, British Columbia, V6T 1Z4,\\ 
Canada\\
Phone: 604 822 2895\\
Fax: 604 822 5485}

\begin{abstract}
Message passing programs commonly use buffers to avoid unnecessary
synchronizations and to improve performance by overlapping
communication with computation.  Unfortunately, using buffers makes
the program no longer portable, potentially unable to complete on
systems without a sufficient number of buffers. Effective buffer use
entails that the minimum number needed for a safe execution be
allocated.

We explore a variety of problems related to buffer allocation for
safe and efficient execution of message passing programs.  We show
that determining the minimum number of buffers or verifying a buffer
assignment are intractable problems.  However, we give a polynomial
time algorithm to determine the minimum number of buffers needed to
allow for asynchronous execution.  We extend these results to several
different buffering schemes, which in some cases make the problems
tractable.

\end{abstract}

\end{frontmatter}

\section{Introduction}

In the last decade MPI~\cite{MPI} and PVM~\cite{PVM} have become the
de facto standards for message passing programs.  They have replaced
the myriad of libraries that provided a degree of portability for
message passing programs.  One aspect of portability introduced in the
MPI standard was that of a {\it safe} program.  As defined in the
standard, a program is safe if it requires no buffering, that is, if
it is synchronous.  Safe programs can be ported to machines with
differing amounts of buffer space.  However, to demand that the
program execute correctly with no buffering is restrictive.  Buffering
reduces the amount of synchronization delay and also makes it possible
to off-load communication to the underlying system or network
components, thus overlapping communication and computation.  Although
one cannot assume an infinite number of buffers, by characterizing the
buffer requirements of a given program it becomes possible to
determine, with respect to buffer availability, whether the program
can be ported to a given machine.  The notion of $k$-safety is
introduced to address the problem of identifying the buffer
requirements of a program to avoid buffer overflows and deadlock.
Determining the minimum $k$, under a variety of buffer placements, is
important for constructing programs that are both safe and can
effectively exploit the underlying hardware.

Unfortunately, the value of $k$ is usually not known a priori.  We
investigate the complexity of determining a minimum value of $k$ for
programs using asynchronous buffered communication with a static
communication pattern and a bounded message size.  We consider the
following three problems: the Buffer Allocation Problem (BAP), which
is the problem of determining the minimum number of buffers required
to ensure deadlock free execution (i.e., determine $k$ for
$k$-safety); the Buffer Sufficiency Problem (BSP), which is to
determine whether a given buffer assignment is sufficient to avoid
deadlock; and finally, the Nonblocking Buffer Allocation Problem
(NBAP),which is to determine the minimum number of buffers needed to
allow for asynchronous execution, that is, when send calls do not
block.

The complexity of these questions also depends on the type of buffers
provided by the system.  We consider four types of system buffering
schemes.  In the first three schemes the buffers are (1) pre-allocated
on the send side only, (2) the receive side only, or (3) mixed and
pre-allocated on both sides.  Finally, we also consider a scheme that
pre-allocates buffers on a per channel basis, where each communication
channel can buffer a fixed number of messages.

We show that the Buffer Allocation Problem is intractable under all
four buffer allocation schemes. The Buffer Sufficiency Problem is
intractable for the receive side buffer and for the mixed buffer
allocation schemes, tractable for the channel scheme and conjectured
tractable for sender side buffers. Finally, the Nonblocking Buffer
Allocation Problem is tractable for all buffer placement schemes,
except the mixed send and receive scheme.

\section{Related Work}

The multiprocess system that we consider is a collection of
simultaneously executing independent asynchronous processes that
compute by interspersing local computation and point-to-point message
passing between processes; these are referred to as {\it
A-computations} in~\cite{CMT96}. Such a system is equivalent to one
with three different events, such as the one defined by
Lamport~\cite{Lamport78}: send events, receive events and internal
events.  As well, we only consider programs that are
repeatable~\cite{CL94a,CL94b} when executed in an unrestricted
environment, that is, programs with static communication patterns.
While this narrows the class of programs we consider, the class of
applications with static communication patterns is still considerable.

The message passing primitives considered in this paper are the
traditional asynchronous, buffered communications: the {\it
nonblocking send} and the {\it blocking receive}, which are the
standard primitives used in MPI and PVM.  Cypher and Leu formally
define the former as a {\it POST-SEND} immediately followed by a {\it
WAIT-FOR-BUFFER-RELEASE} and the latter as a {\it POST-RECEIVE}
immediately followed by a {\it
WAIT-FOR-RECEIVE-TO-BE-MATCHED}~\cite{CL94a,CL94b}.  Informally, the
send blocks until the message is copied out of the process into a send
buffer; the receive blocks until the message has been copied into the
receive buffer.

The notion of safety, as introduced in the MPI standard, underscore
the concern that, when buffer resources are unknown, asynchronous
communication can potentially deadlock the system.  This notion was
extended to $k$-safety, in order to better characterize the buffer
requirements of the program, thus making it safe to take advantage of
asynchronous communication.  The definition of $k$-buffer correctness
was introduced by Bruck~{\it et~al.}~\cite{bruck95} to describe
programs that complete without deadlock in a message passing
environment with $k$ buffers per process.  Similarly, Burns and
Daoud~\cite{burns95ger} introduced {\it guaranteed envelope resources}
into LAM~\cite{burns94}, a public domain version of MPI.  Guaranteed
envelope resources---a weaker condition than $k$-safety---was used in
LAM to reserve a guaranteed number of message header slots on the
receiver side.

Determining whether a system is buffer independent---the system is
0-safe---was investigated in~\cite{CL94a,CL94b}.  In our model, the
interesting systems are buffer-dependent, and require an unknown
number of buffers to avoid deadlock.  

More recently in modern clusters, greater overlap of computation and
communication is possible by off-loading communication onto the
network interface cards.  Unfortunately, most NICs have orders of
magnitude less memory than the average host, which makes message
buffers a limited resource.  Thus, programs that use asynchronous
message passing, and that execute correctly otherwise, might deadlock
when executing on a system where parts of the message passing system
have been off-loaded to the NIC. These issues have been investigated
in several
papers~\cite{MPI,Dongarra93etal,Frye92etal,keppitiyagama2001}.

To determine the minimum number of buffers, the execution of a system
can be modeled using a (coloured) Petri net~\cite{Jensen92}. In order
to determine whether the system can reach a state of deadlock, the Petri
net occurrence graph~\cite{Huber86etal} is constructed, and a search
for dead markings is performed. However, the size of the occurrence
graph is exponential in the size of the original Petri net.

Variations of these problems have been investigated by the operations
research community~\cite{An89,Re87,Sh75}.  In these models, events or
products are buffered between various stations in the production
process, however, the arrival of these events is governed by
probability distributions, which are specified a priori. In our model,
since processes are asynchronous, the time for a message to arrive is
non-deterministic; that is, a message may take an arbitrarily long time
to arrive and a process may take an arbitrarily long time to perform a
send or a receive.

\newcommand{\bs}{\bar{s}}
\newcommand{\br}{\bar{r}}
\newcommand{\bx}{\bar{x}}
\newcommand{\ba}{\bar{a}}
\newcommand{\bb}{\bar{b}}
\newcommand{\bc}{\bar{c}}

\section{Definitions}\label{sec:defs}
Let $S$ be a multiprocess system with $n$ processes and $E_i$
communication events occurring in process $i$; a communication event is
either a send or a receive.  A multiprocess system $S$ is \myem{unsafe}
if a deadlock can occur due to an insufficient number of available
buffers; if $S$ is not unsafe, then $S$ is said to be \myem{safe}.
Figure~\ref{fig:basic} is an example of an unsafe system.  The
numbers above the graph in Figure~\ref{fig:basic} represent the
buffer assignment.

\DoFigure{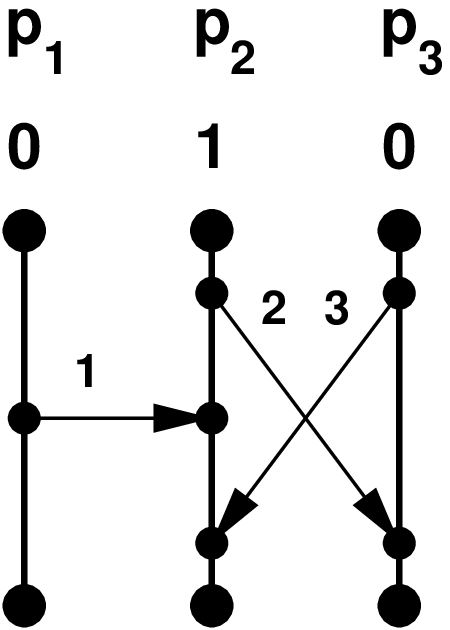}{0.66}{Order of execution can cause deadlock.}{fig:basic}

A per-process \myem{buffer assignment} is an $n$-tuple $B =
(b_1,b_2,\ldots,b_n)$ of non-negative integers representing the number
of buffers that can be allocated by each process.  Similarly, a
per-channel buffer assignment is a $q$-tuple $B =
(b_1,b_2,\ldots,b_q)$, $q =
\Ch{n}{2}$, representing the number of buffers each
channel in the system can allocate.  Since buffers take up memory,
which may be needed by the application, ideally, as few buffers as
possible should be allocated.  However, allocating too few buffers
results in an unsafe system.

Buffer utilization is the nondeterministic phenomena of interest in
the system.  Making the choice of when to use a buffer affects future
choices.  For example, in Figure~\ref{fig:basic}, using a buffer for
communication $1$ before communication 3 completes results in
deadlock.

Two natural decision problems arise from this optimization problem.
Given a system $S$ and a non-negative integer $k$, the \myem{Buffer
Allocation Problem} (BAP) is to decide if there exists a buffer
assignment $B$ such that $S$ is safe and $\sum b_i \leq k$.  In order
to solve this problem we need to solve a simpler one.  Suppose we are
given a buffer assignment $B$ and a system $S$; the \myem{Buffer
Sufficiency Problem} (BSP) is then to decide whether the assignment is
sufficient to make $S$ safe.

Additionally, we can require that no process in system $S$ should ever
block on a send. Given a system $S$ and a non-negative integer $k$, the
\myem{Nonblocking Buffer Allocation Problem} (NBAP) is to decide whether
there exists a buffer assignment $B$, such that no send in $S$ ever
blocks, and $\sum b_i \leq k$.

We model systems by using communication graphs, and executions of
systems by colouring games on these graphs.  Communication graphs can
be derived from execution traces of a program.  The following
subsection defines the graph based framework used throughout this
paper.

\subsection{The Graph Based Framework}
A \myem{communication graph} of $S$ is a directed acyclic graph $G =
G(S) = (V,A)$ where the set of vertices $V = \{v_{i,c}\ |\ 1 \leq i
\leq n, 0 \leq c \leq (E_i+1) \}$ corresponds to the communication
events and the arc set $A$ consists of two disjoint arc sets: the
computation arc set $P$ and the communication arc set $C$.  Each vertex
represents an event in the system: vertex $v_{i,0}$ represents the
\myem{start} of process $i$, vertex $v_{i,c}$, $1 \leq c \leq E_i$,
represents either a \myem{send} or a \myem{receive} event, and vertex
$v_{i,(E_i+1)}$ represents the \myem{end} of a process.  An arc,
$(v_{i,c},v_{i,c+1}) \in P$, $0 \leq c \leq E_i$, represents a
computation within process $i$ and an arc $(v_{i,s},v_{j,t}) \in C$
represents a communication between different processes, $i$ and $j$,
where $v_{i,s}$ is a send vertex, and $v_{j,t}$ is a receive vertex
(e.g.  Figure~\ref{fig:cycle}). Note, the process arcs are drawn
without orientation for clarity; they are always oriented downwards.
Communication graphs are comparable to the time-space
diagrams---without internal events---noted in~\cite{Lamport78}.

\DoFigure{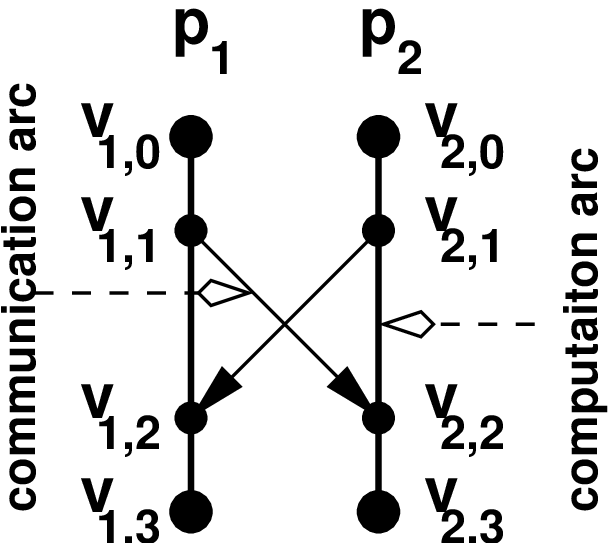}{0.66}{An example of a communication graph with
                                     a 2-ring.}{fig:cycle}

The $i$th \myem{process component} $G_i$ of $G$ is the subgraph
$G_i = (V_i, A_i)$ where $V_i = \{v_{i,c} \in V\ |\ 0 \leq c \leq
(E_i+1)\}$ and $A_i = \{(v_{i,c},v_{i,c+1}) \in A\ |\ 0 \leq c \le
E_i\}$.  The process component corresponds to a process in $S$.  We
construct communication graphs by connecting process components with
arcs.  Hence, it is more intuitive to treat a process component as a
chain of send and receive vertices bound by a start and an end vertex.
A channel is represented by a \myem{channel pair} $(G_i,G_j)$ of 
process components. 

A \myem{t-ring} is a subgraph of a communication graph $G(S)$,
consisting of $t > 1$ process components, such that in each of the $t$
process components there is a send vertex $s_{i_j,c_j}$ and a receive
vertex $r_{i_j,d_j}$, $c_j < d_j$, $1 \leq j \leq t$ such that the
arcs $(s_{i_1,c_1},r_{i_t,d_t})$ and
$(s_{i_{j+1},c_{j+1}},r_{i_j,d_j})$, $1
\leq j < t$ are in $A$. This definition is equivalent to the
definition of a {\it crown} in~\cite{CMT96}.

A t-ring represents a circular dependence of alternating send and
receive events; see the example in Figure~\ref{fig:t-ring}.  The
shaded arcs in Figure~\ref{fig:t-ring} show how each receive event
depends on the preceding send event and each send event depends on the
corresponding receive event. Thus, without an available buffer, there
is a circular dependency that results in the system deadlocking.

\DoFigure{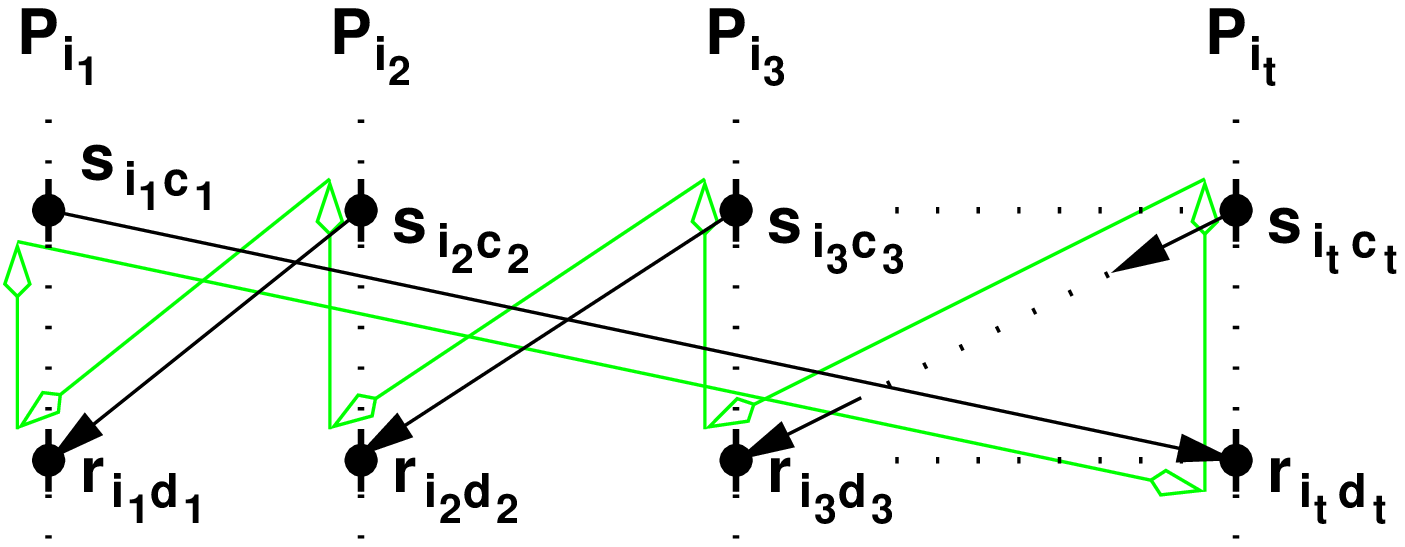}{0.66}{Dependency cycle in $G(S)$.}{fig:t-ring}

To model the execution of a system $S$, we define a colouring game that
simulates the execution of the system with respect to $G(S)$.

\subsection{Colouring the Communication Graph}
Given a communication graph $G(S)$, an execution of a corresponding
system $S$ is represented by a colouring game where the goal is to
colour all vertices green; a green vertex corresponds to the
completion of an event.  We use three colours to denote the state of
each event in the system: a red vertex indicates that the
corresponding event has not yet started, a yellow vertex indicates that
the corresponding event has started but not completed, and a green
vertex indicates that the corresponding event has completed.  Hence, a
red vertex must first be coloured yellow before it can be coloured
green; this corresponds to a traffic light changing from red, to
yellow, to green.\footnote{Naturally, we refer to a European traffic 
light.}  

We use tokens to represent buffer allocations. A buffer assignment of
a process (or channel) is represented by a pool of tokens associated
with the corresponding process component (respectively, the channel
component).  A instance of buffer utilization is represented by
removing a token from a token pool and placing it on the corresponding
communication arc.

The colouring game represents an execution via the following rules.
Initially, the start vertices of $G$ are coloured green and all
remaining vertices are coloured red; this is called the \myem{initial
colouring}.

\newcommand{\SendYellow}{{\it send$\rightarrow$yel}}
\newcommand{\SendGreen}{{\it send$\rightarrow$grn}}
\newcommand{\RecvYellow}{{\it recv$\rightarrow$yel}}
\newcommand{\UseToken}{{\it recv$\stackrel{\bullet}{\rightarrow}$yel}}
\newcommand{\RecvGreen}{{\it recv$\rightarrow$grn}}
\newcommand{\EndYellow}{{\it end$\rightarrow$yel}}
\newcommand{\EndGreen}{{\it end$\rightarrow$grn}}

\begin{tabular}{llp{10.8cm}}
\SendYellow & & A red send vertex may be coloured yellow if the 
                preceding vertex is green---the send is ready.\\
\RecvYellow & & A red receive vertex may be coloured yellow if the  
                corresponding send vertex is yellow, and the preceding 
                vertex (in the same process component) is green---both
                the send and the receive are ready. \\
\UseToken   & & A red receive vertex may be coloured yellow if the 
                corresponding send vertex is yellow, and a token 
                from the corresponding token pool is placed on the 
                incident communication arc---the send is ready and 
                a buffer is used.\\
\SendGreen  & & A yellow send vertex may be coloured green if the 
                corresponding receive vertex is coloured yellow---the 
                communication has completed from the sender's perspective. \\
\RecvGreen  & & A yellow receive vertex may be coloured green if both 
                of its preceding vertices are green.  If the incident 
                communication arc has a token, the token is returned to
                its token pool---a receive completes after the send 
                completes. \\
\EndYellow  & & A red end vertex may be coloured yellow if the preceding 
                vertex is green. \\
\EndGreen   & & A yellow end vertex may be coloured green. 
\end{tabular}

Buffer utilization is represented by placing a token from the token
pool on the selected arc, and colouring the corresponding receive
vertex yellow. If no tokens are available, the rule cannot be invoked.

A \myem{colouring} of $G$, denoted by $\chi$, is a colour assignment
to all vertices, which can be obtained by repeatedly applying the
colouring rules, starting from the initial colouring.  A
\myem{colouring sequence} $\Sigma = (\chi_1,\chi_2,...)$ is a sequence
of colourings such that each colouring is derived from the preceding
one by a single application of one of the colouring rules.  An
execution of a multiprocess system $S$ with buffer assignment $B$ is
represented by a colouring sequence on $G(S)$.  Each transition, from
one colouring to the next, within a colouring sequence, corresponds to
a change of state of an event in the corresponding execution.
Assuming that all events in the system are ordered, there is a
correspondence between the colouring sequences on $G(S)$ and the
executions of system $S$.  Using the correspondence between colouring
sequences on $G(S)$ and executions of system $S$, we reason about
system $S$ by reasoning about colouring sequences on $G(S)$.

We say that a colouring sequence \myem{completes} if and only if the
last colouring in the sequence comprises only green vertices.  A
colouring sequence \myem{deadlocks} if and only if the last colouring
in the sequence has one or more non-green vertices and the sequence
cannot be extended via the application of the colouring rules.
A system $S$ is safe if and only if every colouring sequence on the 
graph $G(S)$ completes.  

We say that a colouring sequence \myem{blocks} if there exists a
sequence on $G(S)$, ending with a colouring containing a yellow send
vertex, that cannot be extended by applying rule \UseToken\ to the
corresponding receive vertex.  A colouring sequence is \myem{block
free} if every prefix of the sequence does not block; a communication
graph $G$, is block free if all colouring sequences on it are also
block free.  If $G(S))$ is block free, then no send in $S$ will ever
block during an execution.

A \myem{token assignment}, also denoted by $B$, is a list of
nonnegative integers, denoting the number of tokens assigned to each
token pool; the token assignment is the abstract representation of a
buffer assignment.  The number of tokens required depends on the
number of times that rule \UseToken\ can be invoked.  If a token pool
is empty, this means all buffers are in use.

\section{Useful Lemmas}
The following lemmas are used throughout our proofs.  Lemma
\ref{lem:t-cycle} characterizes the conditions under which a colouring
sequence will deadlock.  Lemma~\ref{lem:nobuf} characterizes conditions
under which a single colouring sequence may represent all possible
colouring sequences.  Finally, Lemma~\ref{lem:easy} characterizes a class
of communication graphs on which no colouring sequence will deadlock.

\begin{lemma}[The t-Ring Lemma]\label{lem:t-cycle}
Let $G$ be a communication graph comprising a single t-ring.  Any
colouring sequence on $G$ completes if and only if rule \UseToken\ 
is invoked at least once.
\end{lemma}
\begin{proof}
Assume by contradiction that there exists a complete colouring
sequence $\Sigma$ that does not make use of rule \UseToken.  Consider
the first colouring in $\Sigma$ where one of the send vertices is
green; call the vertex $s_i$.  Let $r_j$ be the corresponding receive
vertex.  According to rule \SendGreen, the vertex $r_j$ must be
yellow.  Since rule \UseToken\ has not been applied, rule \RecvYellow\
must have been invoked earlier in the sequence.  By the definition of
a t-ring, the send vertex $s_j$ must be the predecessor of $r_j$.
Since the rule \RecvYellow\ was applied to $r_j$, $s_j$ must be green.
Hence, there is an earlier colouring in $\Sigma$ with a green send
vertex.  This is a contradiction.

In the other direction, if rule \UseToken\ is invoked on receive vertex
$r_j$, then rule \SendGreen\ can be invoked on the corresponding send
vertex $s_j$, breaking the circular dependency.  
\end{proof}

Define the dependency graph of communication graph $G = (V,A)$ to
be $H = (V,E)$ where all process arcs in $A$ are reversed in $E$
and all communication arcs in $A$ are bidirectional in $E$.  Define
the depth $d(v)$ of a vertex $v \in V$ to be the maximum length
path in $H$ from $v$ to a start vertex.

\begin{lemma}\label{lem:nobuf}
Let $G$ be communication graph with a token assignment of $0$.  For any 
vertex $v$ in $G$, if there exists a colouring sequence that colours 
vertex $v$ green, there does not exist a colouring sequence that 
deadlocks before colouring $v$ green.
\end{lemma}
\begin{proof}
Proof by contradiction.  Assume that there exist two colouring
sequences, such that one colouring sequence colours a vertex green and
the other deadlocks and does not colour the vertex green.  Let $v \in
V$ be such a vertex of minimum depth; that is, all vertices of lesser
depth will be coloured green eventually by any colouring sequence on
$G$.  In order for a vertex to be coloured green, its component
predecessor must be green.  Since the depth of the predecessor is less
than the depth of $v$, it can always be coloured green.  Furthermore,
since a send and its corresponding receive vertex are adjacent to each
other, their depths differ by at most $1$.

Since $v$ must be a communication vertex, by rules~\SendGreen\
and~\RecvGreen, the adjacent communication vertex $t$ must be coloured
yellow before $v$ can be coloured green.  If vertex $t$ is of a lesser
depth than $v$, then $t$ must be green colourable in all colouring
sequences; hence, $v$ must also be green colourable.  If $t$ is at the
same depth as $v$, then its component predecessor is at a lesser depth
and must be green colourable, hence $t$ is yellow colourable, and $v$
is green colourable.  If $t$ is at a greater depth than $v$, the
component predecessor of $t$, say $u$, is at the same or a lesser
depth than $t$.  If the latter, then $u$ is green colourable and $t$
is yellow colourable, otherwise, we apply the same argument to $u$
first.  Since there is no path from $u$ to $v$ in $H$---because $d(u)
\leq d(v)$---we need only recurse a finite number of times.
\end{proof}

\begin{lemma}\label{lem:easy}
If $G$ is a communication graph whose dependency graph is acyclic, then
no colouring sequence on $G$ will deadlock.
\end{lemma}
\begin{proof}
Proof by contradiction.  Assume that a colouring sequence deadlocks on
$G$.  Let $v$ be the vertex of minimum depth that cannot be coloured
green.  If $v$ is a send (receive) vertex, let $u$ be the corresponding
receive (send) vertex.  Let vertex $t$ be the component predecessor of
vertex $u$ and let vertex $w$ be the component predecessor of vertex
$v$.  Since the dependency graph is acyclic, the depths of both $t$ and
$w$ are less than the depths of $u$ and $v$.  Hence, both $t$ and $w$
may be coloured green based on our minimality assumption.  However, then
both $u$ and $v$ may be coloured green; this is a contradiction!  If
$v$ is an end vertex, then it has only one predecessor, which is of a 
lesser depth, which leads to the same contradiction.
\end{proof}

\section{Buffer Allocation in Systems with Receive Side Buffers}\label{sec:rbuf}

In systems with receive side buffers, messages are buffered only by the
receiver.  Buffers are allocated by the receiving process when a message
arrives, but cannot be received, and are freed when the message is
received by the application.  Analogously, when colouring a receive
vertex of the corresponding communication graph, only a token belonging
to the same process component may be used.  We call this the
\myem{receive side allocation scheme}.

\subsection{The Buffer Allocation Problem}
In order to prevent deadlock in distributed applications, the underlying
system needs to allocate a sufficient number of buffers.  Ideally, it
should be the minimum number required.  Unfortunately, determining the
required number of buffers, such that the system is safe, is intractable.

The corresponding graph-based decision problem is this: given a
communication graph $G$ and a positive integer $k$, determine if there
is a token assignment of size $k$ such that no colouring sequence
deadlocks on $G$.  We show that \BAPr is $\NPclass$-hard by a
reduction of the well known 3SAT problem~\cite{Co71} to \BAPrn.\
Recall the definition of 3SAT: determine if there exists a satisfying
assignment to $\bigwedge_{i=1}^n (a_i \vee b_i \vee c_i)$, where
$a_i$, $b_i$, and $c_i$ are Boolean literals in
$\{x_1,\bx_1,x_2,\bx_2,\ldots, x_n,\bx_n\}$.

\begin{theorem}\label{thm:bap}
The Buffer Allocation Problem (\BAPrn) is $\NPclass$-hard.
\end{theorem}
\begin{proof}
Proof by reduction of 3SAT to \BAPrn.\  For any 3SAT instance $F$ we
construct a corresponding communication graph $G$ such that for a token
assignment of size $n$, any colouring sequence completes on $G$ if
and only if the corresponding variable assignment satisfies $F$.

Let $F$ be an instance of 3SAT with $n$ variables and $c$ clauses; the
variables are denoted $x_1,x_2,\ldots,x_n$, and the $j$th clause is
denoted $(a_j\vee b_j\vee c_j)$, where $a_j,b_j,c_j \in \{x_1,\bx_1,
\ldots, x_n,\bx_n\}$.  The corresponding communication graph $G$
comprises $2n+1$ process components: $2n$ of the components---called
\myem{literal} components---are labeled $P_{x_i}$ and $P_{\bx_i}$, $i =
1\ldots n$, and correspond to the literals of $F$.  The last
component---called the \myem{barrier} component---is labeled
$P_\mathrm{barrier}$.

Each process component is divided into $c+1$ epochs, where each epoch
is a consecutive sequence of zero or more vertices within the
component.  All epochs are synchronized, that is, the vertices of one
epoch must be coloured green before any of the vertices in the next
epoch may be coloured.  To ensure this we use a barrier component; the
$j$th epoch of the barrier component, $j = 0,\ldots,c$, comprises $2n$
receive vertices, labeled $q_{l,j}$, and $2n$ send vertices, labeled
$t_{l,j}$, $l \in \{x_1,\bx_1,\ldots,x_n,\bx_n\}$.  At the end of each
epoch there is an arc from each of the literal components $P_l$, $l \in
\{x_1,\bx_1,\ldots,x_n,\bx_n\}$, to the barrier component.  Each arc
emanates from vertex $s_{l,j}$, called a barrier send vertex, and is
incident on vertex $q_{l,j}$, where $l \in \{x_1,\bx_1,\ldots,
x_n,\bx_n\}$ and $j = 0\ldots c$.  These arcs are followed by arcs
emanating from the barrier component to the literal components; the
arcs emanate from vertices $t_{l,j}$ and are incident on vertices
$r_{l,j}$, called barrier receive vertices.  The barrier widget has no
cyclic dependencies.  Hence, by Lemma~\ref{lem:easy}, no colouring
sequence will deadlock on a barrier widget.

Epoch $0$ fixes a token assignment corresponding to a variable
assignment in 3SAT.  Each pair of process components, $P_{x_i}$ and
$P_{\bx_i}$, $i = 1\ldots n$, forms a variable widget, which
corresponds to a variable. The two process components of a pair share
a 2-ring; see Figure~\ref{fig:bapr}.  By Lemma~\ref{lem:t-cycle}, at
least one token must be assigned to either process component $P_{x_i}$
or $P_{\bx_i}$ to prevent all colouring sequences from deadlocking on
$G$.  Since only $n$ tokens are available, each component pair can be
assigned exactly one token.  Finally, assigning the token to process
component, $P_{x_i}$ or $P_{\bx_i}$, corresponds to fixing variable
$x_i$ to true or false.  The epoch terminates with a barrier send
vertex $s_{l_i,0}$, followed by a barrier receive vertex $r_{l_i,0}$,
$l_i \in \{x_i,\bx_i\}$.

\DoFigure{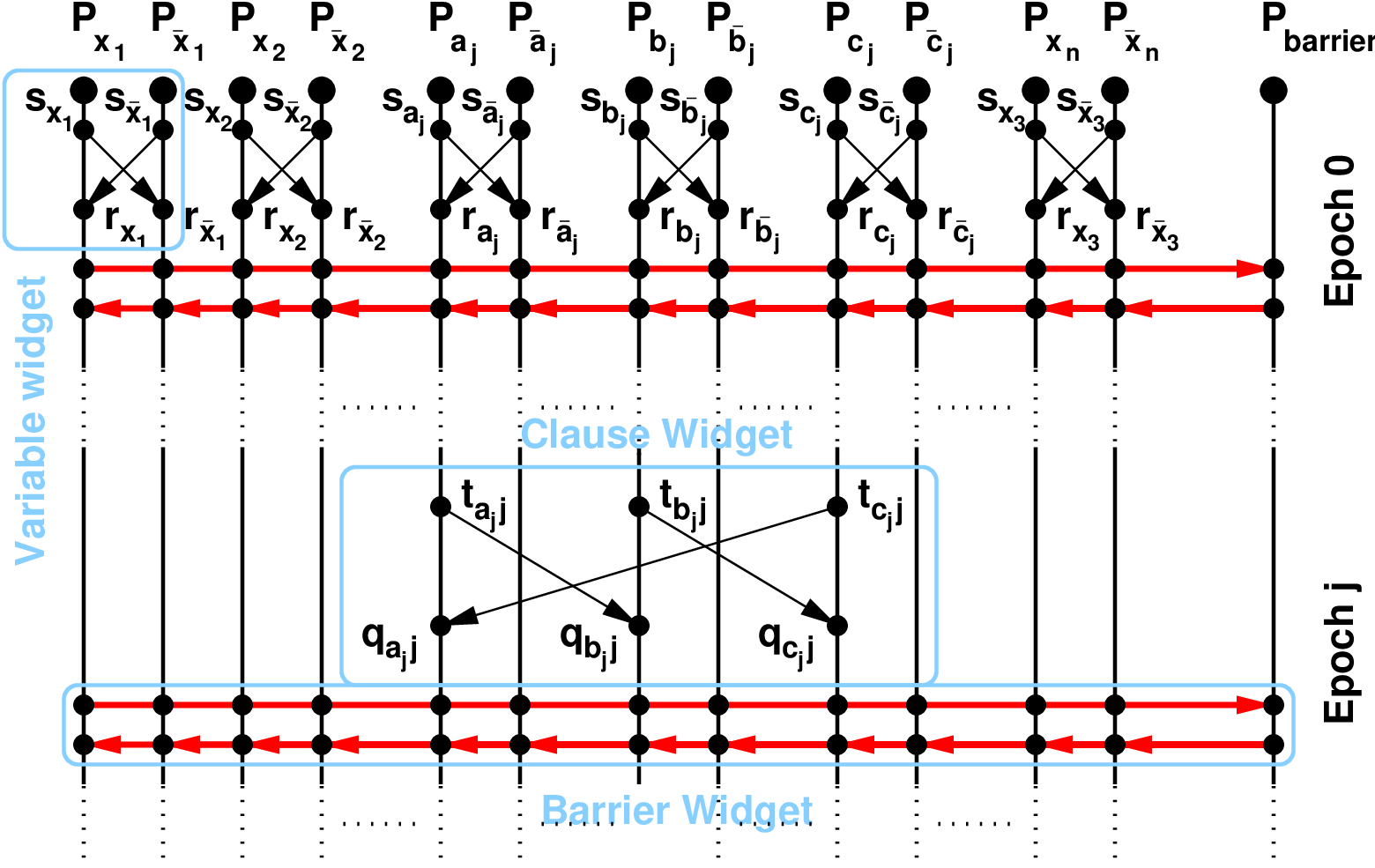}{0.60}{Construction of G.}{fig:bapr}
%\DoDiFigure{figure/fixwid}{0.66}{Construction of epoch $0$.}{fig:fixwid}
%           {figure/clauswid}{0.66}{Construction of epoch $j$.}{fig:clauswid}

Epoch $j$ of each process component corresponds to the $j$th clause of
$F$.  The epoch of a process component $P_l$, $l \not= a_j, b_j,
c_j$---not labeled by a literal of the $j$th clause---contains only
two vertices: the barrier send vertex $s_{l,j}$ and the barrier
receive vertex $r_{l,j}$.  The three process components, $P_{a_j}$,
$P_{b_j}$, $P_{c_j}$, whose labels correspond to the literals in the
$j$th clause share a 3-ring in the $j$th epoch; see
Figure~\ref{fig:bapr}.  By Lemma~\ref{lem:t-cycle}, to avoid deadlock,
at least one of the three process components must have a token. If
none of the components are assigned a token, all literals in the $j$th
clause are false.  The epoch is terminated by the barrier send and the
barrier receive vertices.

A satisfying assignment on $F$ satisfies at least one literal in every
clause.  A corresponding token assignment assigns a token to the
corresponding process component in each 3-ring---corresponding to the
true literal.  Hence, by Lemma~\ref{lem:t-cycle} none of the colouring
sequences will deadlock on any of the clause widgets and any colouring
sequence on $G$ will complete.

For a falsifying assignment of $F$, there exists at least one clause
comprising false literals.  The corresponding token assignment fails
to assign any tokens to the process components that share the
corresponding 3-ring.  Thus, by Lemma~\ref{lem:t-cycle} all colouring
sequences will deadlock in that clause widget.

Hence, for a token assignment of size $n$, any colouring sequence on
$G$ will complete if and only if the corresponding assignment
satisfies $F$.  Since finding a token assignment of size $n$ such that
no colouring sequence on $G$ deadlocks is as hard as finding a
satisfying assignment for $F$, \BAPr is $\NPclass$-hard.
\end{proof}

\subsection{The Buffer Sufficiency Problem}
A potentially simpler problem involved verifying whether a given buffer
assignment is sufficient to prevent deadlock.  Formally, given a graph
$G$ and a token assignment on $G$, determine if none of the colouring
sequences on $G$ deadlock.  This problem turns out to be intractable
as well.

We show that \BSPr is $\coNPclass$-complete by a reduction from the
TAUTOLOGY problem~\cite[Page 261]{GaJo79} to \BSPrn.  Given an instance
of a formula in disjunctive normal form (DNF), $\bigvee^t_{i=1}
\bigwedge_{j=1}^{l_i} a_{i,j}$ where $a_{i,j} \in \{x_1,\bx_1,
\ldots,x_n,\bx_n\}$, the formula is a tautology if it is satisfied by
all assignments.  An assignment that falsifies $F$ is a concise proof
that the formula is not a tautology.  We shall restrict our attention
to 3DNF formulas, where each term has three literals:  $\bigvee^t_{i=1}
(a_i\wedge b_i\wedge c_i)$.

\begin{theorem}\label{thm:bsp}
The Buffer Sufficiency Problem (\BSPrn) is $\coNPclass$-complete.
\end{theorem}
\begin{proof}
Let $G$ be a communication graph along with a token assignment.  If
there exists a deadlocking colouring sequence on $G$, then the sequence
itself is a certificate.  The sequence is at most twice the number of
vertices in $G$.  Hence, \BSPr is in $\coNPclass$.

Let $F$ be a 3DNF formula with $t$ terms where each term has three
literals.  For any 3DNF formula $F$, we construct a communication
graph $G$ and fix a token assignment such that there is a colouring
sequence on $G$ that deadlocks if and only if the corresponding
assignment falsifies $F$.  The construction consists of four types of
widgets that correspond to fixing an assignment, a term in the
disjunction, the disjunction, and the interconnects between widgets.

Each variable in $F$ is represented by a variable widget comprising
three process components that are labeled $P_{x_i}$, $P_{\bx_i}$, and
$Q_i$.  The latter, called the \myem{arbitrator} component, comprises
three receive vertices, labeled $q_i$, $r_{x_i}$, and $r_{\bx_i}$.
The former two process components, called \myem{variable} components,
comprise two send vertices each.  The first, labeled $s_{x_i}$
($s_{\bx_i}$), is adjacent to the corresponding receive vertex
$r_{x_i}$ ($r_{\bx_i})$ in the arbitrator component.  The second,
labeled $t_{x_i}$ ($t_{\bx_i}$), is adjacent to receive vertices in
widgets called dispersers, described later.  The vertex
$q_i$ in the arbitrator component is similarly adjacent to a vertex in
a disperser widget.  The corresponding token assignment for each
variable widget assigns one token to $Q_i$ and no tokens to the other
two components; see Figure~\ref{fig:bspr}.  The widget has the
following property:

\DoFigure{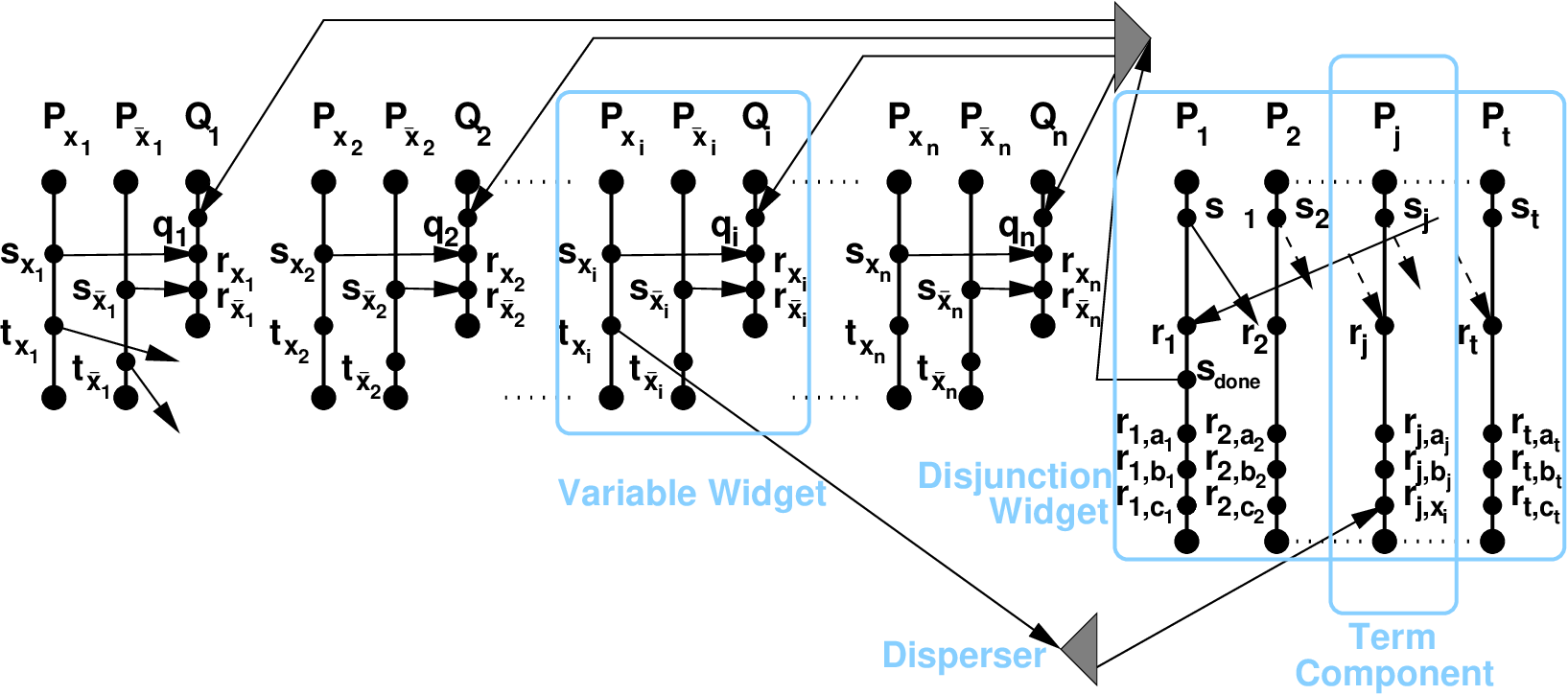}{0.50}{The construction.}{fig:bspr}

\begin{property}\label{prop:variable}
Let $G$ be a communication graph that contains a variable widget.  Any
colouring sequence on $G$ may colour exactly one of the two vertices
$t_{x_i}$ or $t_{\bx_i}$ yellow before vertex $q_i$ is coloured green.
\end{property}
\begin{proof}
By rule \SendYellow, in order for $t_{x_i}$ ($t_{\bx_i}$) to be
coloured yellow, vertex $s_{x_i}$ ($s_{\bx_i}$) must be coloured
green.  Hence, by rule \SendGreen, vertex $r_{x_i}$ ($r_{\bx_i}$) must
first be coloured yellow.  Since vertex $q_i$ is red, vertex $r_{x_i}$
($r_{\bx_i}$) can only be coloured yellow via rule \UseToken.
However, there is only one token assigned to process component $Q_i$,
hence rule \UseToken\ may only be invoked once.
\end{proof}

The $j$th term in the disjunction is represented by a term widget
comprising a process component, which is called the \myem{term}
component and labeled $P_j$.  The first part of each term component
consists of a send vertex $s_j$ and a receive vertex $r_j$; these
vertices are part of a $t$-ring. In the first term component, $P_1$,
there is an additional send vertex labeled $s_\mathrm{done}$; these are
described in the next paragraph.  The second part of each term
component consists of three receive vertices labeled $r_{j,a_j}$,
$r_{j,b_j}$, and $r_{j,c_j}$, where $a_j, b_j, c_j \in \{x_1,\bx_1,
\ldots,x_n,\bx_n\}$ correspond to the literals in the $j$th term; see
Figure~\ref{fig:bspr}.  These receive vertices are adjacent to send
vertices in widgets called dispersers, which are described later.  The
term components are used to construct a disjunction widget.

The disjunction widget comprises $t$ term components, where the first
two vertices, $s_j$ and $r_j$, are part of a $t$-ring spanning all $t$
components.  Specifically, each send vertex $s_j$, $j < t$, is adjacent
to receive vertex $r_{j+1}$ and vertex $s_t$ is adjacent to receive
vertex $r_1$; see Figure~\ref{fig:bspr}.  Each term component is
assigned one token.  The disjunction widget has the following property.

% \DoFigure{figure/sum}{0.66}{The disjunction widget.}{fig:sum}

\begin{property}\label{prop:sum}
Let $G$ be a communication graph that contains a disjunction widget.
Any colouring sequence on $G$ can colour $r_j$, $j\in[1,t]$, green
if and only if at least one of $r_k$, $k\in[1,t]$, is coloured yellow
before any $r_{k,a_k}$, $r_{k,b_k}$, or $r_{k,c_k}$ are coloured
yellow.
\end{property}
\begin{proof}
By Lemma~\ref{lem:t-cycle}, vertex $r_j$ can be coloured green, if and
only if rule \UseToken\ is invoked, colouring one of the receive
vertices $r_k$, $k\in[1,t]$, yellow.  The rule may only be invoked if
and only if a token is available.  Since each term component only has
one token assigned and since vertex $r_k$ precedes vertices $r_{k,a_k}$,
$r_{k,b_k}$, and $r_{k,c_k}$, a token is available if and only if none
of the vertices $r_{k,a_k}$, $r_{k,b_k}$, and $r_{k,c_k}$, are coloured 
yellow via rule \UseToken, before vertex $r_k$ is coloured yellow.  
\end{proof}

Once vertex $r_k$, $k\in[1,t]$, is coloured yellow, all $r_j$,
$j=1\ldots t$ may be coloured green, and vertex $s_\mathrm{done}$ may
be coloured yellow.  We now describe how the widgets are connected
together using disperse widgets.  Let $s$ be a send vertex and
$R$ be a set of receive vertices.  An $(s,R)$-disperser comprises
$|R|+1$ process components: one \myem{master} component, labeled $M_s$,
and $|R|$ \myem{slave} components labeled $S_r$, $r \in R$.  The master
component comprises one receive vertex labeled $r_s$, followed by $|R|$
send vertices labeled $s_r$, $r \in R$.  Each send vertex is adjacent
to the receive vertex on the corresponding slave component $S_r$.  Each
slave component has two vertices: a receive vertex $q_r$, followed by a
send vertex $t_r$; see Figure~\ref{fig:disperser}.  The latter vertex
is adjacent to the receive vertex $r$ in some other widget.  None of
the components are assigned any tokens.  The following property of a
disperser follows from Lemma~\ref{lem:easy}.

\DoFigure{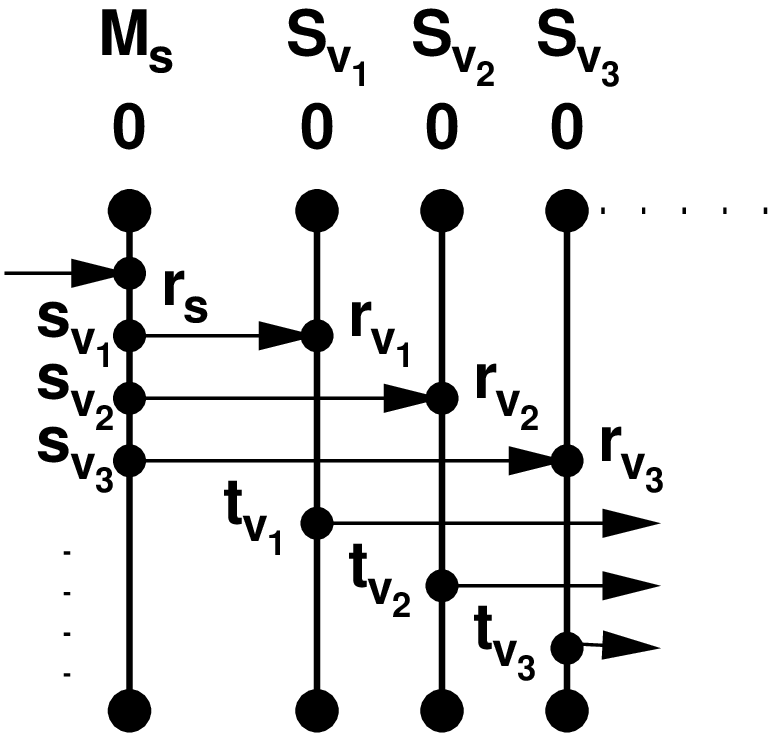}{0.66}{The disperser widget.}{fig:disperser}

\begin{property}\label{prop:disperser}
Let $G$ be a communication graph containing an $(s,R)$-disperser.  If a
colouring sequence colours vertex $r_s$ yellow, then the colouring
sequence can be extended to colour all vertices $t_r$, $r \in R$
yellow.  
\end{property}

Let $R_{x_i}$, $i=1\ldots n$, be the set of receive vertices labeled
$r_{j,x_i} \in P_j$, $j \in [1,t]$, and let $R_{\bx_i}$ be similarly
defined; recall that $a_j, b_j, c_j$ are simply literal place holders
in the vertex labels $r_{j,a_j}$, $r_{j,b_j}$, $r_{j,c_j}$. Hence, a
$(t_{x_i},R_{x_i})$-disperser connects send vertex $t_{x_i} \in
P_{x_i}$ to vertices in $R_{x_i}$---belonging to the term components.
Furthermore, let $Q$ be the set of receive vertices $q_i$ (in the
variable widgets), $i = 1\ldots n$; a $(s_\mathrm{done},Q)$-disperser
connects vertex $s_\mathrm{done}$ to all variable widgets via receive
vertices $q_i$.  The construction of $G$ comprises $n$ variable widgets
and one disjunction widget, composed of $t$ term widgets; these are
connected together by a $(s_\mathrm{done},Q)$-disperser, and $2n$
$(t_a,R_a)$-dispersers, where $a \in \{x_1,\bx_1,\ldots,x_n,\bx_n\}$.
We claim that there exists a colouring sequence that deadlocks on $G$
if and only if there is a falsifying assignment for formula $F$, that
is, $F$ is not a tautology.

Suppose that $F$ has a falsifying assignment $x$, that is every term in
the disjunction is false because each term has a literal $x_i$ or
$\bx_i$, which is false.  To construct a colouring sequence on $G$ that
deadlocks, we construct a set of vertices $U$.  The first half of the
colouring sequence is a maximal colouring sequence involving only the
vertices of $U$.  The second half of the sequence may involve all
vertices in $G$.  The resulting colouring sequence will always
deadlock.

Let $X = \{a \in \{x_1,\bx_1,\ldots,x_n,\bx_n\}\ |\ a|_x = 0\}$, which
is the set of literals that are false, and let $Z = \{s_a \in P_a\ |\ a
\not\in X\} \cup \{s_j\ |\ j=1,\ldots,t\}$, which contains the set of
send vertices from the variable components that are labeled by a true
literal and the numbered send vertices in the disjunction widget; the
set $Z$ contains the vertices which may not initially be coloured.
Let $U = V \backslash Z$ be the rest of the vertex set.

Consider a colouring sequence involving only vertices in $U$.  By
property~\ref{prop:variable} any maximal colouring sequence will
colour the vertices $t_a$ yellow (in the variable widget), where $a
\in X$.  Hence, by property \ref{prop:disperser} the vertices $t_r$
(in the dispersers) will be coloured yellow, where $r \in \bigcup_{a
\in X} R_a$---the send vertices $t_r$ in the dispersers are adjacent
to the receive vertices in $R_a$.  Since $x$ is a falsifying
assignment, every term contains a literal, which is falsified by $x$.
Without loss of generality, let $a_j$ denote a literal that is false
in the $j$th term; therefore, process component $P_j$ contains a
receive vertex $r_{j,a_j}$, which is adjacent to the yellow send
vertex $t_{r_{j,a_j}}$ (in the disperser).  Since none of the vertices
of the $t$-ring (in the disjunction widget) are not in $U$---they are
still coloured red---the token belonging to component $P_j$ is used to
apply rule~\UseToken\ to colour vertex $r_{j,a_j}$ yellow.  Since
every term has a false literal, the colouring sequence colours a
receive vertex $r_{j,a_j}$, $j=1\ldots t$ in every term component
$P_j$.  After the sequence cannot be extended, allow all vertices to
be coloured; since vertices $r_{j,a_j}$ (in the term components),
$j=1\ldots t$, have been coloured yellow before vertex $r_j$ (in term
component $P_j$), according to property \ref{prop:sum}, the sequence
will deadlock.

If a colouring sequence on $G$ deadlocks, according to
property~\ref{prop:sum}, deadlock occurs only if there is a yellow
vertex labeled $r_{k,a_k}$, $r_{k,b_k}$, or $r_{k,c_k}$ in each of the
term components.  Their predecessors---vertices $t_l$, $l \in
\{x_1,\bx_1,\ldots,x_n,\bx_m\}$, in the dispersers---must be green.
Since the colouring sequence is maximal, by
property~\ref{prop:variable} exactly one of $t_{x_i}$ or $t_{\bx_i}$
is red, thus this corresponds to a valid assignment: setting $x_i = 0$
if $t_{x_i}$ is green, or $x_i=1$ if $t_{\bx_i}$ is green yields an
assignment that falsifies $F$.

Thus, a colouring sequence on $G$ deadlocks if and only if the
corresponding assignment falsifies $F$.  Hence, \BSPr is
$\coNPclass$-complete.  
\end{proof}

Therefore, just determining whether a buffer assignment is sufficient
is intractable, even one as simple as in the preceding example.
Intuitively, the buffers of a process are assigned based on the
behaviour of other processes; thus, buffer utilization is not locally
decidable.  Further, the order in which buffers are assigned is
nondeterministic, exploding the search space of possible buffer
utilizations.  This phenomena, which our proofs rely on, is what we
call {\it buffer stealing}.  For example, in a system corresponding to
the variable widget (see Figure~\ref{fig:bspr}), the first process to
send its message gets the buffer, and the other process remains
blocked until the arbitrator performs the receives.  This stealing
corresponds to fixing a value of a variable.  Similarly, the system
corresponding to the disjunction widget allocates buffers for each of
the term processes.  However, if the buffer is stolen in all terms,
corresponding to a falsifying assignment, then the system will
deadlock within the ring.

For completeness, we note the following corollary:
\begin{corollary}
The Buffer Allocation Problem (\BAPrn) is in $\Sigma_2\Pclass$. 
\end{corollary}
\begin{proof}
By Theorem~\ref{thm:bsp}, verifying that a token assignment is
sufficient to prevent deadlock (\BSPrn) is $\coNPclass$-complete.  Since
we can nondeterministically guess a sufficient token assignment, the
result follows.
\end{proof}

\subsection{The Nonblocking Buffer Allocation Problem}\label{sec:nbap}
In addition to the system being safe, we can require that no sending
process ever blocks due to insufficient buffers on the receiving
process.  The Nonblocking Buffer Allocation Problem (\NBAPrn) is to
determine the minimum number of buffers needed to achieve nonblocking
sends.  

Formally, the corresponding decision problem is this: given a communication
graph $G$ and an integer $k$, determine if there exists a token assignment
of size $k$ such that no colouring sequence on $G$ blocks.  Recall
that a colouring sequence does not block if, whenever a send vertex is
coloured yellow, rule \UseToken\ may be applied to the corresponding
receive vertex. 

Let $P_i$ and $P_j$, $j \not= i$, be two process components.  Given
two vertices, $v_{i,c}$ and $v_{i,t}$, in $P_i$, $t > c$, vertex
$v_{i,t}$ is \myem{communication dependent} on vertex $v_{i,c}$ if
$v_{i,c}$ is the start vertex or if there exists a vertex $v_{j,d}
\in P_j$, such that there is a path from $v_{i,c}$ to $v_{j,d}$
and the arc $(v_{j,d},v_{i,t})$ is in $A$ (see Figure~\ref{fig:dep}).
Vertex $v_{i,t}$ is \myem{terminally communication dependent} on
vertex $v_{i,c}$ if $v_{i,t}$ is communication dependent on $v_{i,c}$
and is not communication dependent on the vertices $v_{i,l}$, $c
< l < t$.  The algorithm depicted in Figure~\ref{fig:nbapalg}
computes an optimal token assignment such that no colouring sequence
on $G$ can block.

\DoFigure{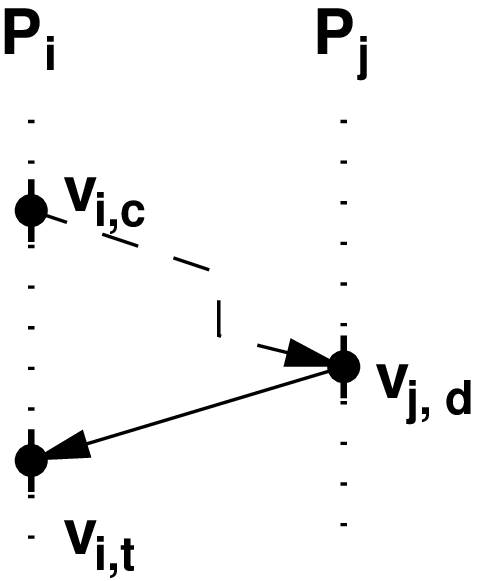}{0.66}{$v_{i,t}$ is communication dependent on 
                                $v_{i,c}$.}{fig:dep}

\begin{figure}[h]
\begin{center}
\fbox{\parbox{5.0in}{
\begin{enumerate}
\setlength{\parskip}{0pt}
\setlength{\partopsep}{0pt}
\setlength{\parsep}{0pt}
\setlength{\itemsep}{0pt}
\item For each receive vertex $v_{i,t}$ determine its terminal 
      communication dependency, vertex $v_{i,c}$, where $t > c$.
\item Set $I_{i,t} = [c,t]$ to be the interval between vertex $v_{i,c}$
      and vertex $v_{i,t}$.
\item For each process component $G_i$, compute $b_i$, the maximum 
      overlap over all intervals $I_{i,t}$.
\item $B = \{b_1,b_2,...,b_n\}$ is the optimal nonblocking token 
      assignment.
\end{enumerate}}}
\end{center}
\caption{Algorithm for computing an optimal nonblocking buffer assignment.
         \label{fig:nbapalg}}
\end{figure}

\begin{remark}
In a system corresponding to communication graph $G$, the time between
a message arriving at process $i$ and its receipt corresponds to the
interval $I_{i,t}$.  Each interval must have a buffer to ensure
nonblocking sends.  Hence, the minimum number of buffers, $b_i$, is the
maximum overlap over all intervals within process $p_i$.  
\end{remark}

Computing the terminal communication dependencies for $G$ can be done
via dynamic programming in $O(|V|n)$ time, where $V$ is the vertex set
of $G$ and $n$ is the number of process components.  If there exists a
path from vertex $v_{i,c}$ to $v_{j,d}$, then there exists a path from
$v_{i,c}$ to all vertices $v_{j,d+k}$, $k > 0$.  Associate with each
vertex $v_{i,c}$ an integer vector $a_{i,c}$ of size $n$; $a_{i,c}[j] =
d$ means that there exists a path from $v_{i,c}$ to $v_{j,d}$, and thus
to $v_{j,d+k}$, $k > 0$.  The vector $a_{i,c}$ is computed by taking
the elementwise minimums over the vectors of the adjacent vertices
$v_{i,c}$; this is simply a depth-first traversal of $G$.  Since the
number of arcs is bounded by $3|V|/2$ and the pairwise comparison takes
$n$ steps, the traversal takes $O(|V|n)$ time.

Next, computing the $O(|V|)$ intervals, $I_{i,t}$, requires one table
lookup per interval.  To compute the maximum overlap we sort the
intervals and perform a sweep, keeping track of the current and maximum
overlap; this takes $O(|V|\log{|V|})$ time.  Thus, the total complexity
is $O(|V|n + |V|\log{|V|})$ time.  In the worst case, where $p \approx 
|V|$, this algorithm is quadratic.  However, in practice $n$ is usually
fixed, in which case the $|V|\log|V|$ term dominates.

\subsubsection{Proof of Correctness of the Nonblocking Buffer Allocation 
               Algorithm}\label{sec:nbapproof}

\begin{lemma}\label{lem:cycle}
Let $G$ be a communication graph. For all vertices $v_{i,c}, v_{j,d} \in 
G$; if $v_{j,d}$ is a send vertex and there exists a path from the vertex 
$v_{i,c}$ to vertex $v_{j,d}$, then vertex $v_{j,d}$ cannot be coloured 
yellow until vertex $v_{i,c}$ is coloured green.
\end{lemma}
\begin{proof}
By rule \SendYellow, the predecessor of $v_{j,d}$ must first be
coloured green before $v_{j,d}$ can be coloured yellow.  Since
rules~\SendGreen, and~\RecvGreen imply that the predecessors
of a green vertex must be green, the result follows.
\end{proof}

\begin{corollary} 
Let $G$, $v_{i,c}$, and $v_{j,d}$ be as in Lemma~\ref{lem:cycle} 
and let $v_{i,t}$ be the receive vertex corresponding to the send 
vertex $v_{j,d}$.  Rule~\UseToken\ will never be applied to vertex 
$v_{i,t}$ before vertex $v_{i,c}$ is coloured green.
\end{corollary}

The preceding corollary implies that a token, which is needed to colour
the receive vertex $v_{i,t}$ yellow, need not be available until the
vertex on which $v_{i,t}$ is terminally communication dependent is
coloured green.  Hence, it is sufficient to ensure token availability
just before colouring the respective send vertex green; this is
also necessary.

\begin{theorem}\label{thm:pebbles}
Given $G$, let $v_{i,c}$ be a send vertex and $v_{i,t}$ be a receive
vertex that is terminally communication dependent on vertex $v_{i,c}$.
A token for the application of rule~\UseToken\ on arc $(v_{j,d},
v_{i,t})$ must be available as soon as vertex $v_{i,c}$ is coloured 
green.
\end{theorem}
\begin{proof}
Let $v_{j,d}$ be the send vertex corresponding to the receive vertex
$v_{i,t}$ and let $Q = \{ v_{i,q} \ |\ c < q < t\}$ be the set of
vertices that are predecessors of $v_{i,t}$, but on which $v_{i,t}$ is
not communication dependent.

Since $v_{i,t}$ is not communication dependent on the vertices in $Q$,
we can construct a colouring sequence on $G$ that fixes the vertices in
$Q$ to be red, and colours vertex $v_{j,d}$ yellow, making the
application of rule \UseToken\ possible in the next step.  Since no
progress is made in the $i$th process component after colouring vertex
$v_{i,c}$ green, the state of the associated token pool does not change
until the application of rule \UseToken\ to vertex $v_{i,t}$.  Hence,
when vertex $v_{i,c}$ is coloured green, the token pool must have a
token destined for arc $(v_{j,d},v_{i,t})$.
\end{proof}

Thus, if a receive vertex $r$ is terminally communication dependent on
a send vertex $s$, then it is necessary and sufficient that a token,
which is used to apply rule \UseToken\ to receive vertex $r$, must be
available as soon as the send vertex $s$ is coloured green; the start
vertex may be thought of as a special send vertex.  Since the interval
corresponding to $r$ begins when $s$ is coloured green, and ends when
$r$ is coloured green, a token must be available for the \UseToken\
rule, which can occur during this interval.  Computing the maximum
overlap of intervals yields the required number of tokens.

\subsubsection{Example Use of the \NBAPr Algorithm}

To demonstrate the \NBAPr algorithm we have implemented it, and
analyzed the pipe-and-roll parallel matrix multiplication
algorithm~\cite{Fox}.  The program has one control process and a
number of worker processes arranged in a 2 dimensional mesh. We ran
the \NBAPr algorithm on meshes of size $2\times 2$, $3\times 3$ and
$4\times 4$.  The communication graph for the smallest example,
comprising four workers ordered in a $2\times 2$ mesh, is depicted in
Figure~\ref{matrix}.  The corresponding optimal buffer assignment is
listed in the second column of Table~1.

\DoFigure{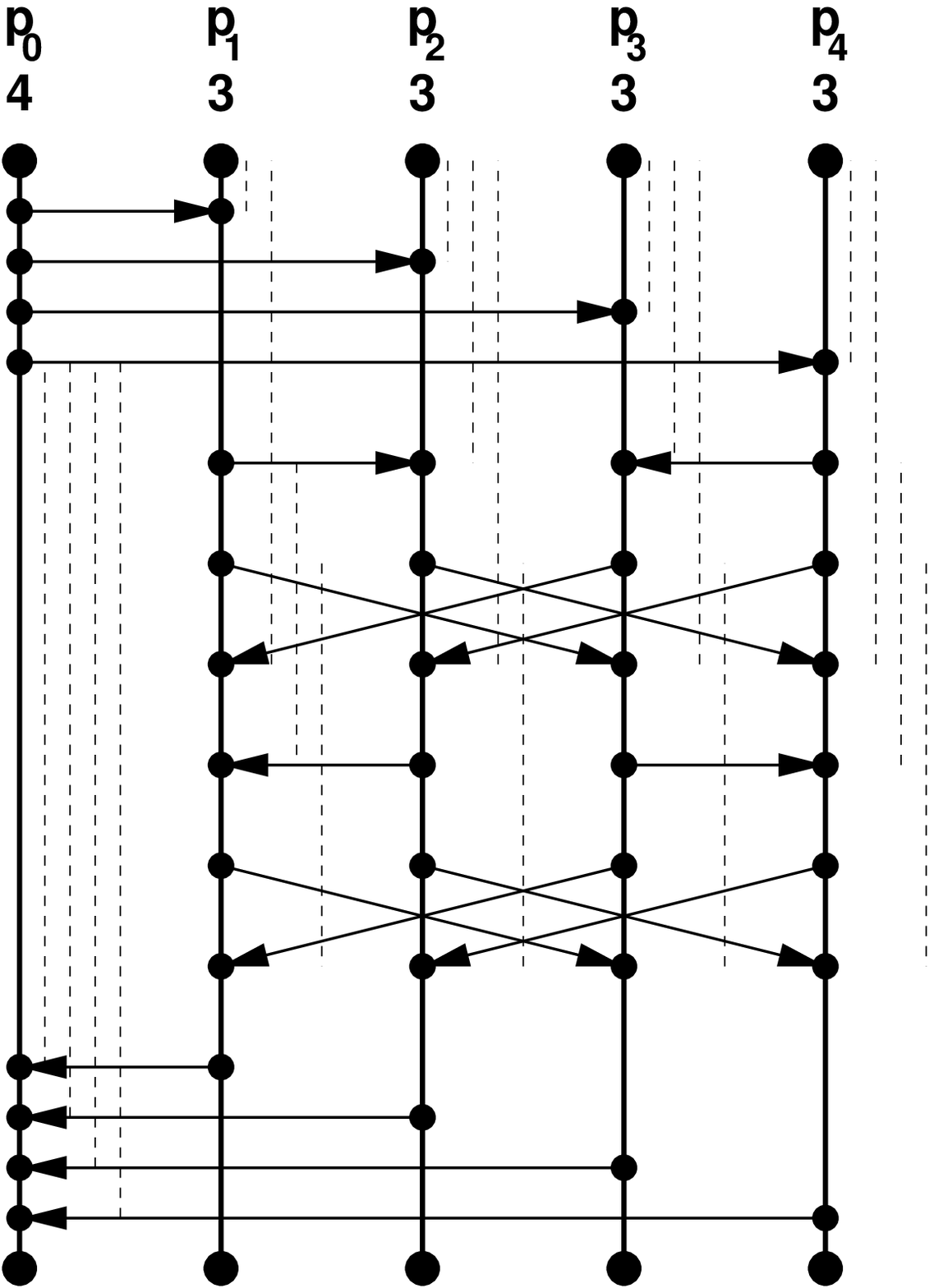}{0.66}{The communication system for a $2\times 2$ worker process 
mesh.}{matrix}

In this example, process 0 is the control process and processes 1
through 4 are the workers. The control process needs 4 buffers and the
workers each need 3 to execute without blocking.  The results obtained
when executing the \NBAPr algorithm on a $3\times 3$ worker system is 9
buffers for the control process and between 4 and 5 buffers for the
worker processes. For the $4\times 4$ system the numbers are 16 for the
control process and between 5 and 7 buffers for the workers.

\begin{table}[h!]
\begin{center}
\begin{small}
\begin{tabular}{|c|c||c|c|c|c|c|c|c|c|c|} \hline
Proc. & Max overlap & \multicolumn{9}{|c|}{Overlap for intervals I$_{\rm j}$}\\ 
& & \multicolumn{1}{|c}{ I$_1$} 
  & \multicolumn{1}{c}{ I$_2$} 
  & \multicolumn{1}{c}{ I$_3$} 
  & \multicolumn{1}{c}{ I$_4$}  
  & \multicolumn{1}{c}{ I$_5$} 
  & \multicolumn{1}{c}{ I$_6$} 
  & \multicolumn{1}{c}{ I$_7$}  
  & \multicolumn{1}{c}{ I$_8$} 
  & \multicolumn{1}{c|}{ I$_9$}\\ \hline\hline
 0    & 4           & 0 & 0 & 0 & 0 & 4 & 3 & 2 & 1 & 0\\
 1    & 3           & 2 & 1 & 2 & 3 & 2 & 1 & 1 & 0 & 0\\
 2    & 3           & 3 & 2 & 1 & 2 & 1 & 1 & 1 & 0 & 0\\
 3    & 3           & 3 & 2 & 1 & 2 & 1 & 1 & 1 & 0 & 0\\
 4    & 3           & 2 & 1 & 2 & 3 & 2 & 1 & 1 & 0 & 0\\ \hline
\end{tabular}
\end{small}\\
{Table 1. The result of running the \NBAPr algorithm on the $2\times2$ worker 
          example.}
\end{center}
\end{table}

\subsubsection{Approximating \BAPr with \NBAPrn}
The \NBAPr algorithm is useful for determining a token assignment that
prevents deadlock, that is, approximating \BAPrn.  Since a nonblocking
colouring sequence does not deadlock, a token assignment determined by
the \NBAPr algorithm ensures that the graph is deadlock free.
However, the token assignment may be far from optimal.  A simple
example of this phenomena is a two process component graph comprised
of $n$ arcs emanating from the first component and incident on the
second.  Such a graph requires zero tokens to avoid deadlock, but
requires $n$ tokens to be block free.  Thus, the aforementioned token
assignment may entail many more tokens than required.

\section{Buffer Allocation in Systems with Send Side Buffers}

In this section we consider the second of the four buffer placement
strategies: send side buffers. Buffers are now allocated on the
sending process side if the receive is not ready to accept the
message.  Correspondingly, the token pool used when applying rule
\UseToken\ to the receive vertex of arc $(s,r)$ belongs to the process
component containing the send vertex $s$.  We call this the \myem{send
side allocation scheme}.

The Buffer Allocation Problem (\BAPsn) remains intractable.  The
problem is conjectured to be $\NPclass$-complete (see the following
paragraph). The $\NPclass$-hardness follows from the observation that
each t-ring in the construction in Theorem~\ref{thm:bap} has to have a
token assigned to a process component pair in order to prevent
deadlock.  It does not matter if the token is allocated from the token
pool of the sending or the receiving process component.  Hence, the
reduction used in Theorem~\ref{thm:bap} can be applied with no
modification.

We conjecture that the corresponding Buffer Sufficiency Problem (\BSPsn)
is in $\Pclass$.  This is because the relative order in which tokens
from a particular token pool are utilized is invariant with respect to
the colouring sequences.  Hence, we believe that the determining
sufficiency is similar to the nonblocking buffer allocation problem and
hence is in $\Pclass$. If this is the case, \BAPs is $\NPclass$-complete.

The Nonblocking Buffer Allocation Problem (\NBAPsn) remains in $\Pclass$.  
The problem can be solved by first reversing all arcs in the communication 
graph, swapping the start and end vertices, and then running the algorithm 
described in Figure~\ref{fig:nbapalg}.

\section{Buffer Allocation in Systems with Send and Receive Side Buffers}

So far we have considered systems exclusively with send side or receive
side buffers.  In this section we investigate systems with buffers on
both the send and the receive sides; many communication systems use
per-host buffer pools for both receiving and sending messages.  The
choice of where to buffer the message---on the sender or on the
receiver---increases the difficulty of determining the system's
properties.

We assume a lazy mechanism for utilizing buffers: first use a buffer
from the sender's pool.  If none is available, use a buffer from the
receiver's pool.  If neither is available, attempt to free a send side
buffer by transferring its contents to a buffer belonging to the
corresponding receiver.  Intuitively, the system attempts to maximize 
buffer use, without attempting to predict the future.  

The corresponding colouring game allows tokens to be allocated from
the pools belonging to both the sending component and the receiving
component.  Correspondingly, a lazy token utilization scheme is used:
let $(s_i,r_j)$ be a communication arc from process component $P_i$ to
process component $P_j$. The following rules apply during the
application of rule \UseToken\ to vertex $r_j$:
\begin{enumerate}
\setlength{\parskip}{0pt}
\setlength{\partopsep}{0pt}
\setlength{\parsep}{0pt}
\setlength{\itemsep}{0pt}
\item If a token belonging to component $P_i$ is available, use it. 
\item Otherwise, if a token belonging to component $P_j$ is available,
      use it.
\item Otherwise, if a token belonging to component $P_i$ is currently
      placed on arc $(t_i,r_k)$, $t_i \in P_i$, $r_k \in P_k$, and a
      token belonging to component $P_k$ is available.  Then the token
      on arc $(t_i,r_k)$ may be replaced with the one belonging to
      $P_k$, freeing a token to be used in the current application of
      rule \UseToken.
\end{enumerate}
We call this the \myem{mixed allocation scheme}.

Not unexpectedly, the Buffer Allocation Problem (\BAPsrn) remains
intractable within the mixed allocation scheme.  This is because
the receive side allocation scheme, which provides no choice of
token pools, can be simulated within the mixed allocation scheme.
Concretely consider the receive side allocation scheme analyzed
in Section~\ref{sec:rbuf}:  to simulate the receive side allocation
scheme on communication graph $G$, within the mixed allocation
scheme, each arc in $G$ is replaced by the widget illustrated in
Figure~\ref{fig:sr2r}.  Since vertex $q$ cannot be coloured green until
vertex $r$ is coloured yellow, and component $\Pp$ has no tokens, applying
rule~\UseToken\ to $r$ requires that $P_j$ has an available token, 
regardless of whether $P_i$ has an available token.

\DoFigure{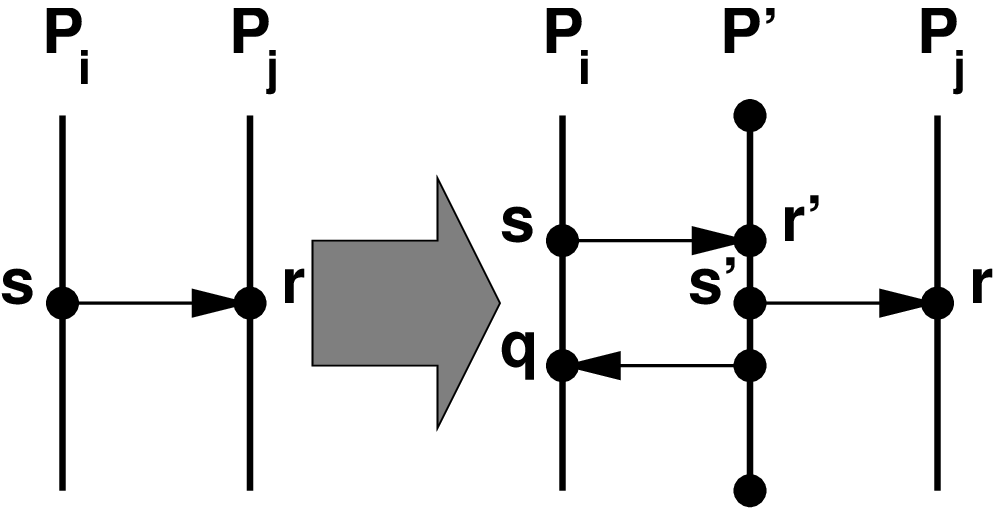}{0.66}{Nullifying send side token pools.}{fig:sr2r}

Similarly, the Buffer Sufficiency Problem (\BSPsrn) within the mixed
allocation scheme is also $\coNPclass$-complete.  The hardness follows
from Theorem~\ref{thm:bsp} and the preceding argument.  Since a
colouring sequence also serves as a deadlock certificate in this case,
the $\coNPclass$-completeness result follows.

The interesting property of the mixed allocation scheme is that
the Nonblocking Buffer Allocation Problem (\NBAPsrn) is intractable;
the choice of token pools increases the search space of solutions
exponentially!  The reduction is from 3SAT.

\begin{theorem}
The Nonblocking Buffer Allocation Problem (\NBAPsrn) is $\NPclass$-hard.
\end{theorem}
\begin{proof}
Let $F$ be an instance of 3SAT, comprising $n$ variables, labeled
$x_i$, $i=1\ldots n$, and $c$ clauses.  We construct a communication
graph $G$ such that there exists a token assignment of $n+2$ tokens
that prevents any colouring sequence from blocking on $G$ if and only
if the corresponding assignment satisfies $F$.

The graph $G$ comprises $2n+3$ process components: the first $2n$ are
labeled $P_{x_i}$ and $P_{\bx_i}$, $i = 1\ldots n$, and the remaining
three process components are labeled $P$, $Q_0$ and $Q_1$,
respectively.  The graph is divided into $c+1$ epochs:  epoch $0$
corresponds to the variable assignment, and epochs $1$ through $c$
correspond to clause evaluation.

In epoch $0$ each process component $P_{x_i}$ contains a single send
vertex $s_i$ that is adjacent to the receive vertex $r_i$ located in
epoch $0$ of process component $P_{\bx_i}$.  Process component $Q_0$
(and $Q_1$) contains four vertices: two receive vertices $q_{0,1}$ and
$q_{0,2}$ (respectively $q_{1,1}$ and $q_{1,2}$), followed by two send
vertices $t_{0,1}$ and $t_{0,2}$ (respectively $t_{1,1}$ and
$t_{1,2}$).  Finally, process component $P$ contains eight vertices:
two send vertices, $s_{0,1}$ and $s_{0,2}$, that are adjacent to
vertices $q_{0,1}$ and $q_{0,2}$; two receive vertices, $r_{0,1}$ and
$r_{0,2}$, that are adjacent to $t_{0,1}$ and $t_{0,2}$; two more send
vertices, $s_{1,1}$ and $s_{1,2}$, that are adjacent to $q_{1,1}$ and
$q_{1,2}$; and two more receive vertices, $r_{1,1}$ and $r_{1,2}$,
that are adjacent to $t_{1,1}$ and $t_{1,2}$.  See
Figure~\ref{fig:nbapsr}.  Epoch $0$ has two important properties.

\begin{property}\label{prop:litbuf}
Any token assignment must assign at least one token to either
component $P_{x_i}$ or $P_{\bx_i}$ to prevent the colouring sequence
from blocking after colouring vertex $s_i$ yellow.
\end{property}

\begin{property}\label{prop:2buf}
A token assignment on $G$ having only $n+2$ tokens must assign two
tokens to process component $P$ to prevent a colouring sequence from
blocking after yellow colouring one of the send vertices $s_{0,1}$,
$s_{0,2}$, $s_{1,1}$ or $s_{1,2}$.
\end{property}
\begin{proof}
Since $n$ tokens must be allocated to the process components $P_{x_i}$
or $P_{\bx_i}$, $i = 1,\ldots,n$, this leaves only two tokens to be
allocated.  Since the colouring rule sequence \SendYellow, \UseToken,
\SendGreen, \SendYellow, \UseToken can colour send vertices $s_{0,1}$
and $s_{0,2}$, or send vertices $s_{1,1}$ and $s_{1,2}$, component
pairs $(P,Q_0)$ and $(P,Q_1)$ must each have two tokens between them.
This can only happen by assigning the tokens to $P$.  
\end{proof}

A corollary of these properties is that once a legal token assignment
is made, no colouring sequence will block in epoch 0.  The choice of 
allocating the token on $P_{x_i}$ versus $P_{\bx_i}$ corresponds to 
fixing the variable assignment.

\DoFigure{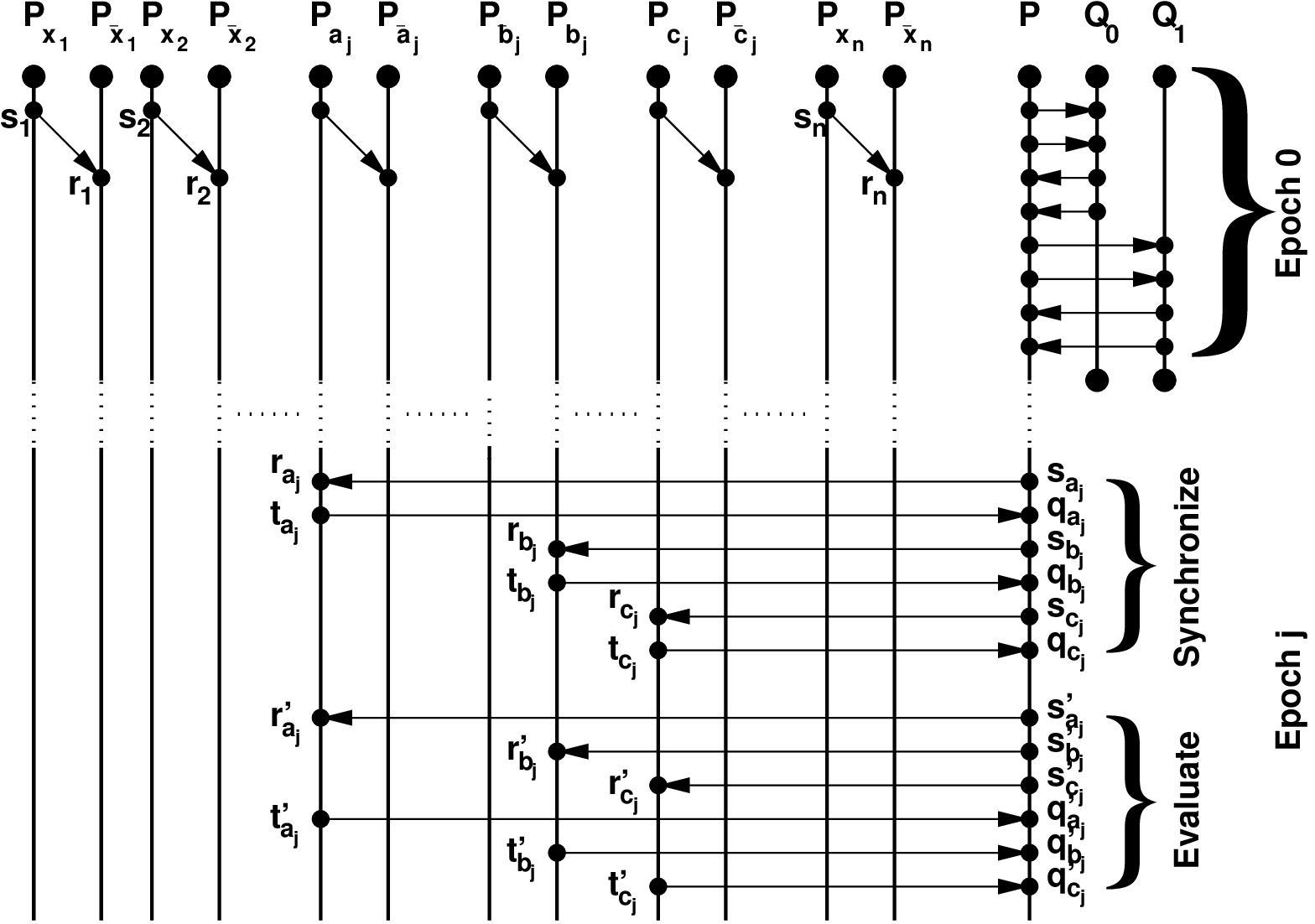}{0.60}{Reduction from 3SAT to \NBAPsr.}{fig:nbapsr}

For $j = 1\ldots c$, epoch $j$ corresponds to the $j$th clause.  Each
epoch comprises two parts of six arcs each: the synchronization part
and the evaluation part.  Four process components are involved in an
epoch: the three components, $P_{a_j}$, $P_{b_j}$, and $P_{c_j}$, whose
labels are the literals in the $j$th clause, where $a_j,b_j,c_j \in
\{x_1,\bx_1,\ldots,x_n,\bx_n\}$, and component $P$, which is involved
in every epoch.  Epoch $j$ of component $P_{a_j}$ comprises four
vertices:  receive vertex $r_{a_j,j}$, send vertex $t_{a_j,j}$, receive
vertex $\rp_{a_j,j}$, and send vertex $\tp_{a_j,j}$.  Process
components $P_{b_j}$ and $P_{c_j}$ are analogously formed.

In epoch $j$ component $P$ has 12 vertices, the first six are these:
send vertex $s_{a_j,j}$, receive vertex $q_{a_j,j}$, send vertex
$s_{b_j,j}$, receive vertex $q_{b_j,j}$, send vertex $s_{c_j,j}$, and
receive vertex $q_{c_j,j}$.  These are followed by three send
vertices: $\spr_{a_j,j}$, $\spr_{b_j,j}$, and $\spr_{c_j,j}$, and
three receive vertices: $\qp_{a_j,j}$, $\qp_{b_j,j}$, and
$\qp_{c_j,j}$.

Each vertex $s_{l,j}$ is adjacent to vertex $r_{l,j}$, each vertex
$t_{l,j}$ is adjacent to vertex $q_{l,j}$, each vertex $\spr_{l,j}$ is
adjacent to vertex $\rp_{l,j}$, and each vertex $\tp_{l,j}$ is
adjacent to vertex $\qp_{l,j}$; see Figure~\ref{fig:nbapsr}.  For
conciseness we drop the last index, $j$, if it is obvious from the
context.  Epoch $j$ has three important properties:

\begin{property}\label{prop:epoch}
If vertex $\qp_{c_j}$ (in epoch $j$) is coloured green and vertex
$s_{a_{j+1}}$ (in epoch $j+1$) is still red, then no tokens that
belong to component $P$ are assigned to arcs.  The same applies to
vertex pairs $(q_{a_j}, s_{b_j})$, $(q_{b_j},s_{c_j})$, and
$(q_{c_j},\spr_{a_j})$, also in epoch $j$.
\end{property}
\begin{proof}
All ancestors of $\qp_{c_j}$ must be coloured green and all descendants
of $s_{a_{j+1}}$ must be coloured red.  This includes all vertices
in $G$, except some vertices $s_i$ and $r_i$ in epoch $0$, which are not
adjacent to vertices in component $P$.  Hence, the tokens belonging to
$P$ are not assigned to any arc.  The same argument applies to the other
vertex pairs.  
\end{proof}

\begin{property}\label{prop:sync}
A colouring sequence on $G$ can block only when yellow colouring receive 
vertices $\rp_{a_j}$, $\rp_{b_j}$, $\rp_{c_j}$, $\qp_{a_j}$, 
$\qp_{b_j}$, or $\qp_{c_j}$.
\end{property}
\begin{proof}
As a corollary of properties~\ref{prop:litbuf} and~\ref{prop:2buf}, no
colouring sequence can block in epoch $0$.  Thus, we need only check
that no colouring sequence can block in the first part of epoch $j$,
$j=1\ldots c$.

By property~\ref{prop:epoch}, if $s_{a_j}$ is red and its predecessor 
is green, then no tokens of $P$ are in use.  Hence, to colour
$s_{a_j}$ green, a token is available to colour $r_{a_j}$ yellow.
Since vertex $r_{a_j}$ is a predecessor of $t_{a_j}$,  vertex
$r_{a_j}$ must be coloured green before $t_{a_j}$ may be coloured
yellow.  Thus the token is freed before $t_{a_j}$ is coloured green,
and may be used to colour vertex $q_{a_j}$ yellow after $t_{a_j}$
is coloured yellow.  A similar argument applies to the vertices
$r_{b_j}$, $q_{b_j}$, $r_{c_j}$, and $q_{c_j}$.  
\end{proof}
\begin{property}\label{prop:three}
A colouring sequence can block in epoch $j$ if and only if none of the
three process components, $P_{a_j}$, $P_{b_j}$, and $P_{c_j}$, have a
token assigned.
\end{property}
\begin{proof}
For the `if' direction consider a colouring sequence that colours
vertex $q_{c_j}$ green, but has not yet coloured vertex $\spr_{a_j}$
yellow.  By definition, blocking does not occur, if rule \UseToken\
may always be applied to colour a receive vertex yellow.  To colour
the send vertices $\spr_{a_j}$, $\spr_{b_j}$, and $\spr_{c_j}$ yellow
and then green, the receive vertices $\rp_{a_j}$, $\rp_{b_j}$, and
$\rp_{c_j}$, must be coloured yellow via rule \UseToken.  Since the
receive vertices $\rp_{a_j}$, $\rp_{b_j}$, and $\rp_{c_j}$ are not
ancestors of the send vertices $\spr_{a_j}$, $\spr_{b_j}$, and
$\spr_{c_j}$, none of the receive vertices need be coloured green
before the send vertices are coloured yellow.  However, component $P$
has only two tokens, and none of components $P_{a_j}$, $P_{b_j}$,
$P_{c_j}$ have any. Hence, rule \UseToken\ can only be invoked twice,
instead of the requisite three times.  Thus, a colouring sequence can
block in epoch $j$.

For the `only if' direction we claim that if a literal component
$P_{a_j}$, $P_{b_j}$, or $P_{c_j}$ has a token, rule \UseToken\ can be
invoked on any of the six receive vertices $\rp_{a_j}$, $\rp_{b_j}$,
$\rp_{c_j}$, $\qp_{a_j}$, $\qp_{b_j}$, and $\qp_{c_j}$.  Since
$\rp_{a_j}$ is a predecessor of $\tp_{a_j}$, $\rp_{a_j}$ must be
coloured green before $\tp_{a_j}$, and hence before $\qp_{a_j}$ is
coloured yellow.  Thus, the same token that was allocated upon the
application of rule \UseToken\ to vertex $\rp_{a_j}$, may also be
allocated upon the application of rule \UseToken\ to vertex
$\qp_{a_j}$; the same argument is applicable to vertices $\qp_{b_j}$
and $\qp_{c_j}$.  Applying rule \UseToken\ to vertices $\rp_{a_j}$ and
$\rp_{b_j}$, uses the two tokens from component $P$.  To colour vertex
$\rp_{c_j}$ yellow there are three possible scenarios:
\begin{enumerate}
\setlength{\parskip}{0pt}
\setlength{\partopsep}{0pt}
\setlength{\parsep}{0pt}
\setlength{\itemsep}{0pt}
\item the colouring sequence has already freed one of the tokens, 
      allowing it to be reused, 
\item component $P_{c_j}$ has a token, in which case it is used, or
\item component $P_{a_j}$ (or $P_{b_j}$) has a token, in which case it 
      replaces the token used to yellow colour vertex $\rp_{a_j}$ 
      (or $\rp_{b_j}$) and the freed token is used to colour vertex 
      $\rp_{c_j}$.
\end{enumerate}
Since at least one component $P_{a_j}$, $P_{b_j}$, or $P_{c_j}$ have a 
token, the claim is proven.
\end{proof}

By property~\ref{prop:three} a colour sequence will block in epoch $j$
if and only if none of the process components $P_{a_j}$, $P_{b_j}$, or
$P_{c_j}$ has a token, which corresponds to the $j$th clause having no
literals that are true.  Thus, a token assignment of size $2n+2$
prevents any colouring sequence on $G$ from blocking if and only if the
corresponding assignment satisfies $F$.  
\end{proof}

\newcommand{\fmv}[2]{\langle#1,\mathrm{#2}\rangle}

\section{Buffer Allocation in Channel Based Systems}
In channel based systems processes communicate via pairwise
connections that are created at start-up.  Each connection, called a
channel, is specified by its end-points and is used by one process to
send messages to the other.  Each channel functions independently of
other channels in the system, and resources such as buffers are
allocated on a per channel basis, rather than per process.  Finally,
channels behave like queues, that is, messages are removed from the
channel in the same order that they are inserted.

Channels may either be unidirectional, comprising source and
destination end-points, or bidirectional, comprising two symmetric
end-points.  In the former case, only the source process may insert
messages into the channel and only the destination process may remove
messages from the channels.  A bidirectional channel is equivalent to
two unidirectional channels, allowing both processes to insert and
remove messages from the channel.  Here we only consider unidirectional
channels.

Except for buffer allocation, channel based communication does not
differ from the previously described send/receive mechanism.  In fact,
an unbuffered channel communication is just a synchronous send/receive
communication.  Thus, we can derive similar results for channel based
systems.

In the corresponding colouring game tokens, are allocated to channels
(component pairs) instead of to components.  This change does not change
the properties used in our proofs.  In fact, Lemma~\ref{lem:t-cycle} may
be used unchanged.  We call this the \myem{per channel allocation
scheme}.

\subsection{The Buffer Allocation Problem}
The corresponding Buffer Allocation Problem (\BAPucn) is this: given a
communication graph $G$ and an integer $k$, determine whether there
exist a token assignment of size $k$, such that no colouring sequence
deadlocks on $G$.  Even though token utilization, during the colouring
of a communication graph, is only dictated by the communication arcs
within a process component pair, determining the number of tokens
needed remains $\NPclass$-hard.  The proof is similar in spirit to
Theorem~\ref{thm:bap}.

\begin{theorem}\label{thm:bap-uc}
The Buffer Allocation Problem (\BAPucn) is $\NPclass$-hard.
\end{theorem}
\begin{proof}
We prove this by reducing 3SAT to \BAPucn.\  For any 3SAT instance $F$
we construct a corresponding communication graph $G$---polynomial in
size of $F$---such that for a token assignment of size $n$, any
colouring sequence will complete on $G$ if and only if the
corresponding variable assignment satisfies $F$.

Let $F$ be an instance of 3SAT on $n$ variables and comprising $c$
clauses.  The construction is nearly identical to that in
Theorem~\ref{thm:bap}, except for the widgets representing the clauses
of $F$.   The graph $G$ has $2n$ process components that are labeled by
the literals of $F$, $P_{x_i}$ and $P_{\bx_i}$, $i=1\ldots n$.  Each
component comprises $c+1$ epochs, where each epoch contains zero or two
vertices.

As in Theorem~\ref{thm:bap}, epoch $0$ fixes a variable assignment.  In
epoch $0$ each component has two vertices: a send vertex, labeled
$s_{x_i}$ (or $s_{\bx_i}$), and a receive vertex $r_{x_i}$,
(respectively $r_{\bx_i}$), $i = 1\ldots n$.  Vertex $s_{x_i}$ is
adjacent to vertex $r_{\bx_i}$, and vertex $s_{\bx_i}$ is adjacent to
vertex $r_{x_i}$; this is a 2-ring, identical to epoch $0$ in
Theorem~\ref{thm:bap}.  Epoch $0$ has the the following property:

\begin{property}\label{prop:valid}
Any colouring sequence on $G$ will deadlock in epoch $0$ unless each
process component pair has a token assigned to the token pool of
either $(P_{x_i},P_{\bx_i})$, or $(P_{\bx_i},P_{x_i})$, $i = 1\ldots
n$. Thus, the token assignment must be of at least size $n$.  (Follows
from Lemma~\ref{lem:t-cycle}.)
\end{property}

Property~\ref{prop:valid} yields the following correspondence between
assignments on $F$ and token assignments of size $n$.

\begin{property}\label{prop:needed}
The corresponding token assignment of a variable assignment on $F$
assigns a token to the channel $(P_{x_i},P_{\bx_i})$ if $x_i$ is
true, or to $(P_{\bx_i},P_{x_i})$ if $x_i$ is false.  
\end{property}

The $j$th epoch represents the $j$th clause of $F$, denoted $(a_j,
b_j, c_j)$, where $a_j, b_j, c_j \in \{x_1,\bx_1,\ldots, x_n,
\bx_n\}$.  The process components $P_{a_j}$, $P_{\ba_j}$, $P_{b_j}$,
$P_{\bb_j}$, $P_{c_j}$, and $P_{\bc_j}$ form a 6-ring, while the
remaining components have no vertices in the $j$th epoch.  Process
component $P_{a_j}$ has two vertices in the $j$th component: a send
vertex, $s_{a_j,j}$, and a receive vertex $r_{a_j,j}$; similarly,
the other five components have a send and receive vertex that are
correspondingly named.  The arcs linking the 6 components are these:
$(s_{a_j,j},r_{\ba_j,j})$, $(s_{\ba_j,j}, r_{b_j,j})$,
$(s_{b_j,j},r_{\bb_j,j})$, $(s_{\bb_j,j}, r_{c_j,j})$,
$(s_{c_j,j},r_{\bc_j,j})$, and $(s_{\bc_j,j},r_{a_j,j})$.  These
form a 6-ring, as illustrated in Figure~\ref{fig:ucclauswid}.  The
key property of the $j$th epoch is this:

\DoFigure{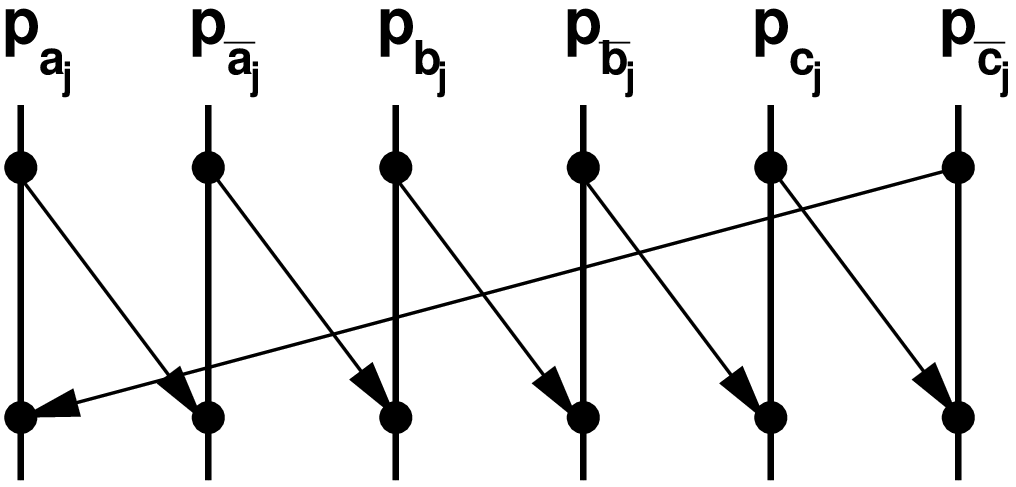}{0.66}{The clause representation in epoch $j$.}{fig:ucclauswid}

\begin{property}\label{prop:ucclaus}
No colouring sequence on $G$ will deadlock in the $j$th epoch if and
only if at least one of the channels has a token:
$(P_{a_j},P_{\ba_j})$, $(P_{\ba_j}, P_{b_j})$, $(P_{b_j},P_{\bb_j})$,
$(P_{\bb_j},P_{c_j})$, $(P_{c_j}, P_{\bc_j})$,
$(P_{\bc_j},P_{a_j})$. (Follows from Lemma~\ref{lem:t-cycle}.)
\end{property}

A refined version of property~\ref{prop:ucclaus} is more useful:

\begin{property}\label{prop:refine}
For any token assignment of size $n$ such that no colouring sequence
deadlocks on $G$ in epoch $0$, no colouring sequence on $G$ will
deadlock in the $j$th epoch if and only if at least one of the
channels $(P_{a_j}, P_{\ba_j})$, $(P_{b_j},P_{\bb_j})$, and
$(P_{c_j},P_{\bc_j})$, has a token.
\end{property}
\begin{proof}
By property~\ref{prop:valid}, all token assignments that do not
cause deadlock in epoch $0$ only assign tokens to channels of the
form $(P_{x_i},P_{\bx_i})$ or $(P_{\bx_i},P_{x_i})$.  Hence, only
channels $(P_{a_j}, P_{\ba_j})$, $(P_{b_j},P_{\bb_j})$, and
$(P_{c_j},P_{\bc_j})$ can have a token.  By property~\ref{prop:ucclaus},
no colouring sequence on $G$ will deadlock in epoch $j$ if one of
these channels has a token.  
\end{proof}

We claim that given a token assignment of size $n$, any colouring sequence
will complete on $G$ if and only if the corresponding variable assignment
satisfies $F$.

If an assignment $x$ satisfies $F$, then every clause has at least one
literal that evaluates to true.  By Property~\ref{prop:needed}, in each
of the $j$ epochs at least one of the channels listed in
Property~\ref{prop:refine} will be allocated a token.  Hence, by
Property~\ref{prop:refine} no colouring sequence will deadlock on $G$.

If an assignment $x$ does not satisfy $F$ then there is at least one
clause in which all literals are false.  Let $(a_j,b_j,c_j)$ be the
unsatisfied clause.  By property~\ref{prop:needed}, the corresponding
token assignment will not assign a token to $(P_{a_j},P_{\ba_j})$,
$(P_{b_j},P_{\bb_j})$, or $(P_{c_j},P_{\bc_j})$, hence, by
Property~\ref{prop:refine}, all colouring sequences will deadlock.

Thus, \NBAPuc is $\NPclass$-hard.
\end{proof}

Since tokens are assigned on a per channel basis, token usage depends
only on the two process components that comprise the channel.
Consequently, the sufficiency of a token assignment can be verified in
linear time.  Thus, the easier problem \BSPuc is in $\Pclass$, implying
that \BAPuc is $\NPclass$-complete.  We describe the verification
algorithm and prove its correctness.

To verify the sufficiency of a token assignment, perform a colouring
on $G$: at each step of the colouring a vertex of $G$ is coloured
according to the rules in section~\ref{sec:defs}.  Using a queue to
keep track of colourable vertices, means that determining which vertex
to colour next takes $O(1)$ time.  Since each vertex changes colour at
most twice---the maximum length of any colouring sequence is $2|V|$
colourings---colouring a graph takes $O(|V|)$ time.  The token
assignment is sufficient if and only if the colouring sequence
completes.  The algorithm's correctness follows immediately from the
following theorem: any colouring sequence on $G$ completes if and only
if some colouring sequence on $G$ completes.  Thus, a token assignment
is sufficient if and only if some colouring sequence on $G$ completes.

\begin{theorem}
Let $G$ be a communication graph and $B$ a token assignment on $G$.  
Any colouring sequence on $G$ completes if and only if a colouring
sequence on $G$ completes.
\end{theorem}
\begin{proof}
For any communication graph $G$, we construct a new graph $\Gp$ where
every token is simulated by a process component, the size of the
corresponding token assignment is zero, and  every colour sequence on
$G$ corresponds to a colouring sequence on $\Gp$, such that a colouring
sequence on $G$ completes if and only if the corresponding colouring
sequence on $\Gp$ completes.  Since the token assignment on $\Gp$ is
zero, by Lemma~\ref{lem:nobuf} a colouring sequence on $\Gp$ completes
if and only if every colouring sequence on $\Gp$ completes.  Hence,
every colouring sequence on $G$ completes if and only if a colouring
sequence on $G$ completes.

To simulate an $m$ token channel (a channel that has been assigned $m$
tokens) $m$ process components are chained together.  For each channel
$(P,Q)$ with $m$ tokens, $m$ process components $P_1,P_2,\ldots,P_m$
are interspersed between $P$ and $Q$.  The channel $(P,Q)$ is replaced
with these channels: $(P,P_1),(P_1,P_2),\ldots,(P_{m-1},P_m),(P_m,Q)$.
Each arc from $P$ to $Q$ is replaced by a chain of arcs from $P \to P_1
\to P_2 \to \ldots \to P_{m-1} \to P_m \to Q$.  The replacement is
illustrated in Figure~\ref{fig:pbuf}.

\DoFigure{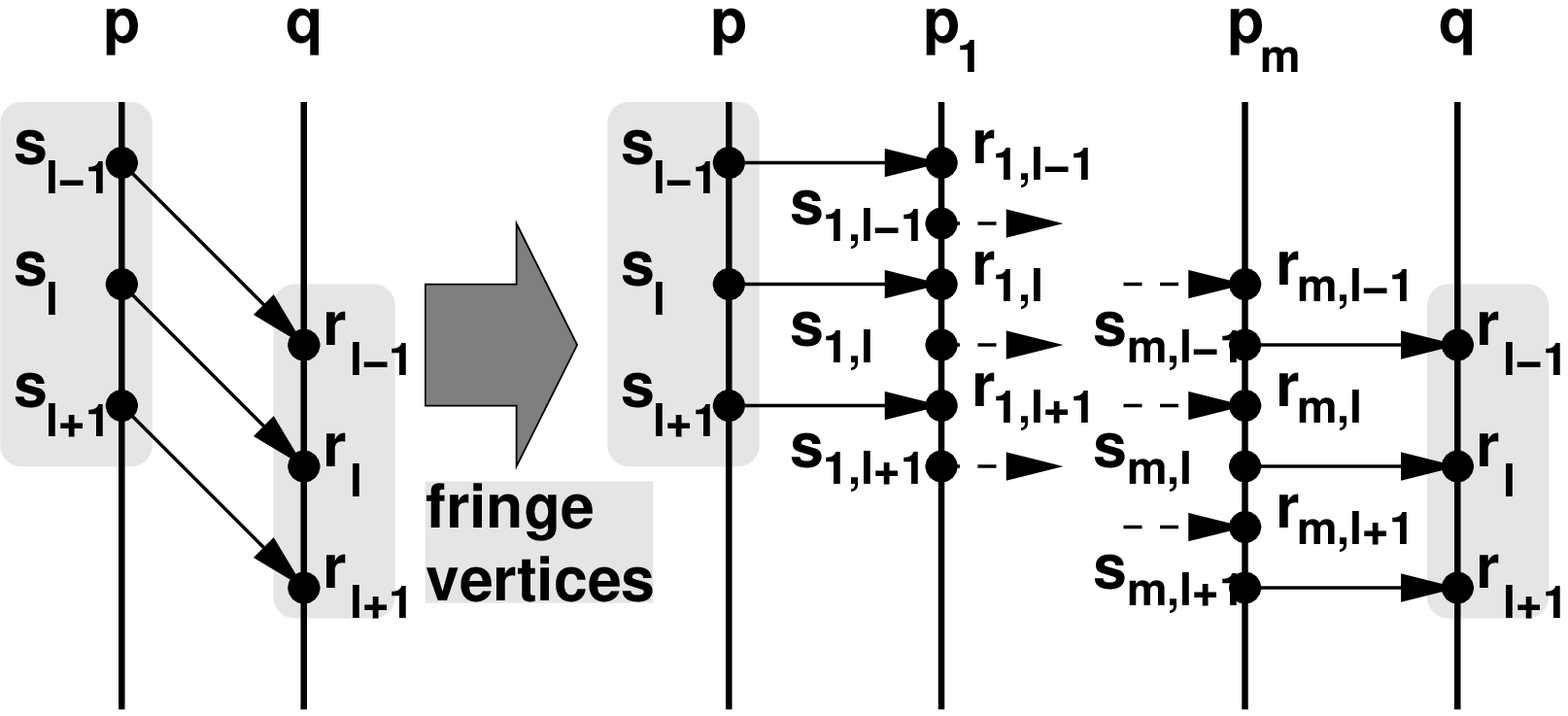}{0.66}{Simulating $m$ tokens by $m$ components.}{fig:pbuf}

We claim that a colouring sequence, $\Sigma$, on $G$ will deadlock if
and only if the corresponding colouring sequence, $\Sigmap$, on $\Gp$
deadlocks.  First, we construct the correspondence and argue its
correctness.  Second, we argue that sequence $\Sigma$ deadlocks on $G$
if and only if the corresponding sequence $\Sigmap$ deadlocks on $\Gp$.
Finally, we apply Lemma~\ref{lem:nobuf} to prove our result.

Since the transformation is iterative---each $m$ token channel is 
independent of the other channels---it is sufficient to derive the 
correspondence between the colouring sequence on $G$ and the graph $\Gp$ 
where a single $m$ token channel has been replaced.  Let $(P,Q)$ denote
the channel in $G$ that is replaced in $\Gp$.

Let $(s_l,r_l) \in G$, $l = 1,2,\ldots$, denote the arcs from
process component $P$ to $Q$.  The corresponding paths in $\Gp$ are
\[ (s_l,\underbrace{r_{1,l},s_{1,l}}_{P_1},\underbrace{r_{2,l},s_{2,l}}_{P_2},
  \ldots, \underbrace{r_{m,l},s_{m,l}}_{P_m},r_l), \]
where each arc $(r_{k,l},s_{k,l})$ is within process component $P_k$
and each arc $(s_{k,l},r_{k+1,l})$ is between process components $P_k$
and $P_{k+1}$; the vertices $s_l$ and $r_l$, $l = 1,2,\ldots$ are
called the \myem{fringe} vertices.

A colouring sequence $\Sigma$ can be represented as a sequence of
differences (or moves), $\delta_i$, between every two consecutive
colourings $\chi_i$ and $\chi_{i+1}$.  The sequence $\Delta_\Sigma =
\delta_1\delta_2\ldots$ is a sequence of colouring game moves $\delta_i
= \fmv{v}{colour}$ such that applying $\delta_i$ to colouring $\chi_i$
yields $\chi_{i+1}$, the next colouring in $\Sigma$; $\Delta_\Sigma$
can be derived from $\Sigma$ and, $\Sigma$ can be derived from
$\Delta_\Sigma$ and $G$.  The sequence $\Delta_\Sigma$ comprises two
types of moves: those that colour fringe vertices, called \myem{fringe
moves}, and those that do not, called \myem{normal moves}.

Given a colouring sequence $\Sigma$ on $G$, we transform it into
the corresponding colouring sequence $\Sigmap$ on $\Gp$.  The
transformation replaces some fringe moves in sequence $\Delta_\Sigma$
with sequences of moves, resulting in the corresponding move sequence
$\Delta_\Sigmap$.  This sequence comprises normal moves and
\myem{added moves}; added moves are a mixture of fringe moves and
moves on the vertices within the added components $P_i$. There are
four types of fringe moves in $\Delta_\Sigma$: colour a send vertex
$s_l$ yellow ($\fmv{s_l}{yel}$), colour a send vertex $s_l$ green
($\fmv{s_l}{grn}$), colour a receive vertex $r_l$ yellow
($\fmv{r_l}{yel}$), and colour a receive vertex $r_l$ green
($\fmv{r_l}{grn}$).  The transformation is performed in the order
that the moves occur in sequence $\Delta_\Sigma$.

\begin{itemize}
\setlength{\parskip}{0pt}
\setlength{\partopsep}{0pt}
\setlength{\parsep}{0pt}
\setlength{\itemsep}{0pt}
\item If $\delta_i = \fmv{s_l}{yel}$, then no action is taken.

\item If $\delta_i = \fmv{s_l}{grn}$, we replace it with the sequence 
\[\fmv{r_{1,l}}{yel},\fmv{s_l}{grn},\fmv{r_{1,l}}{grn},
  \fmv{s_{1,l}}{yel},\]
suffixed by the sequences
\[\fmv{r_{j,l}}{yel},\fmv{s_{j-1,l}}{grn},\fmv{r_{j,l}}{grn},
  \fmv{s_{j,l}}{yel},\ \ j = 2 \ldots k-1\]
where $k$ is the smallest integer such that the move
$\fmv{s_{k,l-1}}{grn}$ has not yet been inserted into the move sequence
$\Delta_\Sigma$, that is, vertex $s_{k,l-1}$ has not yet been coloured
green.  

\item If $\delta_i = \fmv{r_l}{yel}$, we remove it from the sequence; 
it is restored when we replace the move $\fmv{r_l}{grn}$.  

\item If $\delta_i = \fmv{r_l}{grn}$  we replace this move with the sequence
\[\fmv{r_l}{yel},\fmv{s_{m,l}}{grn},\fmv{r_l}{grn},\]
suffixed with the sequences
\[\fmv{r_{g_j,h_j}}{yel},\fmv{s_{g_j-1,h_j}}{grn},
  \fmv{r_{g_j,h_j}}{grn}, \fmv{s_{g_j,h_j+1}}{yel},\ \ j=0\ldots k-1,\]
where $g_j = m - j$, $h_j = l + 1 + j$, and $k$ is the smallest integer
such that the move $\fmv{s_{m-k,l+1+k}}{yel}$ has not yet been inserted
into the sequence, that is, vertex $s_{m-k,l+1+k}$ has not yet been
coloured yellow.  Since the head of this sequence colours $s_{m,l}$
green, $r_{m,l+1}$ could be coloured yellow, if $s_{m-1,l+1}$ is
yellow, then $s_{m-1,l+1}$ could be coloured green followed by
$r_{m,l+1}$ and finally $s_{m,l+1}$ could be coloured yellow; this
colouring cascades down the added process components.  
\end{itemize}
It is important to note that each of the replacement sequences is maximal,
that is, no additional valid colouring moves on the chain process
components $P_i$, $i = 1\ldots m$, may be suffixed to them.  The new
sequence looks like this:
\[\Delta_\Sigmap = 
\overbrace{\delta_1\ldots\delta_{h_1}}^\mathrm{normal\ moves}
\underbrace{\deltap_1\ldots\deltap_{g_1}}_\mathrm{added\ moves}
\overbrace{\delta_{h_1+1}\ldots\delta_{h_2}}^\mathrm{normal\ moves}
\underbrace{\deltap_{g_1+1}\ldots\deltap_{g_2}}_\mathrm{added\ moves} \ldots.\]

Since $G$ is a contraction of $\Gp$, all normal vertices are coloured
by $\Delta_\Sigmap$ in the same order as in $\Delta_\Sigma$.  Recall
that normal vertices are not adjacent to the process component chain,
and hence, are not affected by the transformation.  While normal
vertices within process components $P$ and $Q$ may depend on the order
that the fringe vertices are coloured, the dependence is via process
arcs, not communication arcs.  Consequently, the normal vertices only
depend on the order that the fringe vertices are coloured green.
Fortunately, this order is preserved.  By inspection, the replacement
sequences of moves are valid.  Thus, the transformed sequence
$\Delta_\Sigmap$ is valid.   Additionally, all green colouring moves on
fringe vertices are preserved by the transformation; a vertex is
coloured green by $\Delta_\Sigma$ if and only if the corresponding
vertex is coloured green by $\Delta_\Sigmap$.  The following property
is key:

\begin{property}\label{prop:key}
$\Delta_\Sigma$ deadlocks on $G$ if and only if $\Delta_\Sigmap$ 
deadlocks on $\Gp$.
\end{property}
\begin{proof}
By contradiction, suppose that $\Delta_\Sigma$ deadlocks on $G$ while
$\Delta_\Sigmap$ can be extended, that is, another vertex colouring
move may be suffixed to $\Delta_\Sigmap$.  Let $v$ be the vertex that
can be coloured by the extension.  Vertex $v$ may either be a normal
vertex, a fringe vertex, or a vertex belonging to a process chain.  The
latter is impossible because every replacement sequence of moves is
maximal.

If $v$ is a normal vertex, then its predecessors are either a normal
vertex or a fringe vertex that has been coloured green.  Since the
transformation preserves the colourings of normal vertices and the
order in which vertices are coloured green, if $\Delta_\Sigmap$ can be
extended by colouring $v$, then so can $\Delta_\Sigma$, which is a
contradiction.

If $v$ is a fringe vertex, there are four cases: either $v$ is a send
vertex $s_l$ being coloured yellow or green, or $v$ is a receive
vertex $r_l$ being coloured yellow or green.  The transformation does
not affect moves that colour send vertices yellow and such a colouring
only depends on its process component predecessor being green.  Hence,
if the colouring can be suffixed to $\Delta_\Sigmap$, it can also be
suffixed to $\Delta_\Sigma$; resulting in a contradiction.  If the
extension colours the send vertex green, this means that the original
sequence $\Delta_\Sigma$ can be extended by either adding the
colourings $\fmv{s_l}{grn}$ or $\fmv{r_l}{yel}\fmv{s_l}{grn}$,
depending on whether $r_l$ has been coloured yellow or not in the
original sequence $\Delta_\Sigma$; thus, it is a contradiction.

Similarly, if $v$ is a fringe receive vertex being coloured green, this
is not possible, because the transformation colours fringe receive
vertices yellow, then green, by a single replacement sequence.
Finally, if $v$ is a fringe receive vertex $r_l$ that can be coloured
yellow, the original sequence $\Delta_\Sigma$ can be extended by the
move $\fmv{r_l}{grn}$, because in the original sequence the
corresponding send vertex $s_l$ has already been coloured green. Thus,
we have another contradiction.

In the other direction, if the original sequence can be extended, then
transforming the extension of the sequence $\Delta_\Sigma$ yields an
extension to the presumably deadlocked sequence $\Delta_\Sigmap$.
Thus, $\Delta_\Sigma$ deadlocks on $G$ if and only if $\Delta_\Sigmap$
deadlocks on $\Gp$.  
\end{proof}

A corollary of Property~\ref{prop:key} is that the colouring sequence 
$\Sigma$ deadlocks if and only if the colouring sequence $\Sigmap$ 
deadlocks.

By Lemma~\ref{lem:nobuf} a colouring sequence on $\Gp$ completes if
and only if all colouring sequences on $\Gp$ complete.  Hence, a
colouring sequence on $G$ completes if and only if all colouring
sequences on $G$ complete.
\end{proof}

\begin{corollary}
A colouring sequence on $G$ completes if and only if the token assignment 
is sufficient.
\end{corollary}

\subsection{The Buffer Allocation Problem}
For the Nonblocking Buffer Allocation Problem, the algorithm derived in
section~\ref{sec:nbap} suffices with a small modification.  Since the
token pools are per channel, rather than per process component, the
computation must be performed on a per pool basis. Hence, there is an
additional factor of $n$ in the runtime.  Since each process may be
using up to $n$ channels, the runtime of the algorithm becomes
$O(|V|n^2 + |V|n\log{(|V|n)})$; the cost increases because the number
of allocations to be made becomes quadratic in $n$.

\section{Conclusion}

As message passing becomes increasingly popular, the problem of
determining $k$-safety plays an increasingly important role. The
relevance of this problem is grows as more and more functionality of
message passing systems is off-loaded to the network interface card,
where limited buffer space is a serious issue. Even if message passing
is kept in main memory, buffer space can still be limited due to the
sometimes very large data sets used in many parallel and distributed
programs. Unfortunately, determining $k$-safety is intractable.

We have shown that in the receive buffer model, determining the number
of buffers needed to assure safe execution of a program is
$\NPclass$-hard, and that even verifying whether a number of assigned
buffers is sufficient is $\coNPclass$-complete. On the positive side,
if we require that no send blocks, we provide a polynomial time
algorithm for computing the minimum number of buffers.  By allocating
this number of buffers, safe execution is guaranteed. In addition, we
have implemented the \NBAPr algorithm, and it is now part of the {\sf
Millipede} debugging system~\cite{Pe03}.

For systems with only send buffers, the Buffer Allocation Problem
remains $\NPclass$-complete. In addition, we conjecture that the
Buffer Sufficiency Problem can be solved in polynomial time because
the order of the sends in each process is fixed. The Nonblocking
Buffer Allocation problem for systems with only send buffers can be
solved in polynomial time.

For systems with both send and receive buffers, the Buffer Allocation
Problem as well as the Buffer Sufficiency Problem remain
intractable. More interestingly, the Nonblocking Buffer Allocation
problem has become intractable.

For systems with unidirectional channel buffers, both the Buffer
Sufficiency Problem as well as the Nonblocking Buffer Allocation
Problem have polynomial time algorithms. However, the Buffer
Allocation Problem still remains an $\NPclass$-complete problem.
The results (conjectures) are summarized below.

\begin{center}
\begin{tabular}{|l||c|c|c|c|} \hline
& \multicolumn{4}{|c|}{Buffer Placement}\\
Problem & Receive & Send & Send \& Receive & Channel\\\hline\hline
\myem{BAP} & $\NPclass$-hard       & $\NPclass$-hard & 
$\NPclass$-hard       & $\NPclass$-complete \\\hline
\myem{BSP} & $\coNPclass$-complete & ($\Pclass$)     &
$\coNPclass$-complete & $\Pclass$           \\ \hline
\myem{NBAP}& $\Pclass$             & $\Pclass$       &
 $\NPclass$-hard      & $\Pclass$           \\ \hline
\end{tabular}
\end{center}

\subsection{Strategies for Reducing Buffer Requirements}

There are several strategies that a programmer can use to reduce the
likelihood of deadlock when only a few buffers are available.  

The obvious solution is to use synchronous communication, which does
not require any buffers at all. However, this is not always desirable.

For efficiency reasons asynchronously buffered communication is often
preferred.  To decrease the risk of deadlock the programmer can
introduce epochs that are separated by barrier synchronizations.  This
might reduce the number of buffers needed for each epoch, as no buffer
requirement spans an epoch boundary. If each epoch only needs a small
number of buffers, the risk of deadlock due to buffer insufficiency is
reduced.

\bibliographystyle{plain}
\bibliography{ksafe}
\end{document}